\documentclass[prd,twocolumn,floatfix,nofootinbib]{revtex4}

\usepackage{epsfig} 
\usepackage{amsmath} 
\usepackage{amsfonts}

\newcommand{\beq}{\begin{equation}}
\newcommand{\eeq}{\end{equation}}
\newcommand{\beqa}{\begin{eqnarray}}
\newcommand{\eeqa}{\end{eqnarray}}

\newcommand{\ome}{\Omega_e}
\newcommand{\ceasy}{CMBeasy} 
\newcommand{\lcdm}{$\Lambda$CDM}
\newcommand{\lmax}{l_{\rm max}}
\newcommand{\dlss}{d_{\rm lss}}
\newcommand{\mnu}{\sum m_\nu} 
\newcommand{\Lv}{\mathbf{L}}
\newcommand{\lv}{\mathbf{l}}
\newcommand{\xv}{\mathbf{x}}

\begin{document} 

\title{CMB Lensing Constraints on Neutrinos and Dark Energy} 
\author{Roland de Putter, Oliver Zahn, Eric V.\ Linder}
\affiliation{Berkeley Lab \& Berkeley Center for Cosmological Physics, 
University of California, Berkeley, CA 94720}  

\date{\today}

\begin{abstract} 
Signatures of lensing of the cosmic microwave background radiation by 
gravitational potentials along the line of sight carry with them 
information on the matter distribution, neutrino masses, and dark 
energy properties.  We examine the constraints that Planck, PolarBear, 
and CMBpol future data, including from the B-mode polarization 
or the lensing potential, will be able to place on these quantities. 
We simultaneously fit for neutrino mass and dark energy equation 
of state including time variation and early dark energy density, and 
compare the use of polarization power spectra with an optimal quadratic 
estimator of the lensing.  Results are given as a function of systematics 
level from residual foreground contamination.  A realistic CMBpol 
experiment can effectively 
constrain the sum of neutrino masses to within 0.05 eV and the 
fraction of early dark energy to 0.002.  We also present a surprisingly 
simple prescription for calculating dark energy equation of 
state constraints in combination with supernova distances from JDEM. 
\end{abstract} 

\maketitle

\section{Introduction \label{sec:intro}}

Precision studies of the cosmic microwave background (CMB) have helped us 
formulate a standard model of cosmology and measure several global 
parameters that describe our universe and its contents 
\cite{Komatsu:2008hk,acbar,Ruhl:2002cz,Readhead:2004gy}.  
Six key parameters to describe the cosmology 
have been determined with 1-10\% precision and CMB data plays a 
significant role in constraining other parameters, such as spatial 
curvature, the dark energy density, and the Hubble constant, in 
combination with other types of data. 

However, we know in some cases and allow the possibility in other 
cases, that there are further fundamental parameters beyond the six. 
One example is the mass of neutrinos, where terrestrial experiments 
indicate a nonzero, though unknown, value: $m_{\nu} \gtrsim 0.05$ eV 
for at least one neutrino species \cite{numass}. 
Another set of parameters of great interest describes the properties 
of the dark energy causing acceleration of the cosmic expansion.  
The dark energy equation of state (EOS) may differ from the constant 
value $w=-1$ of the cosmological constant, and may vary with time. 
Indeed, this dynamics would be a key clue to the nature of the 
physics behind acceleration.  The persistence of dark energy density 
to early times is another mystery that is crucial to explore.  Current 
CMB data on the temperature and E-mode polarization spectra (and their 
cross-spectra) are of little use in themselves in addressing these 
issues, and this holds to a large extent even in combination with 
other cosmological information such as supernova distances and large 
scale structure data. 

Fortunately, other types of CMB information exist, though they have 
not yet been measured.  This includes the CMB deflection field -- 
the action of gravitational potentials along the line of sight on the 
CMB -- and the B-mode polarization spectra (and cross-spectra) resulting 
from this.   The effects carry contributions from all redshifts between 
the source (the last scattering surface at redshift $z\approx1090$) and 
the observer, though like all lensing deflection the kernel from the 
geometric distance factors peaks approximately midway, $z\approx3-4$. 
The growth of the gravitational potentials over this history carries 
within it information on the matter power spectrum.  Thus the effects 
of neutrino masses and dark energy properties are encoded in the CMB. 

While subsets of these effects have been investigated before (see, e.g., 
\cite{KapKnowSong03,scr,lesg,Smith:2006nk,cmbpollens}), the effects have not generally 
been considered simultaneously (especially for dynamical dark energy), 
with the critical covariances between them.  This article is also the first 
investigation of the important question of early dark density using 
CMB lensing.  We also examine for a range of cases the added leverage 
of lensing information extraction through use of the optimal quadratic 
estimator which utilizes the unique non-Gaussian structure in the map 
caused by lensing. 

In \S\ref{sec:method} we lay out the methodology for obtaining precision 
theoretical predictions for power spectra, and their slight variations 
with cosmology, and summarize the observational capabilities of three 
benchmark CMB surveys.  We explore adding neutrino mass to the standard, 
cosmological constant universe in \S\ref{sec:lcdm}, and include as 
well the dark energy EOS and its time variation in \S\ref{sec:wa}. 
Discussion includes complementarity with other cosmological probes 
and issues of foreground noise.  In \S\ref{sec:ede} we investigate early 
dark energy density, and present a simple prescription for cosmological 
constraints in \S\ref{sec:jdem}.  We summarize the key prospects for 
intermediate range CMB experiments in \S\ref{sec:inter}.

\section{Power Spectra Modeling: Theory and Experiments \label{sec:method}} 

Primordial perturbations in the photon number density arise from 
Gaussian, random, adiabatic fluctuations seeded in the inflationary 
era.  These induce a photon temperature power spectrum, and interaction 
with inflationary gravitational waves and scattering from electrons 
creates $B$-mode and $E$-mode polarization power spectra (as well as 
a TE cross-spectrum), respectively.  Gravitational lensing shuffles the 
photon pattern on the sky \cite{lin89,Seljak:1995ve} and contributes to 
each of these spectra, as well as transforming some of the $E$-modes 
into $B$-modes, introducing a coupling between the two.  Beyond these 
power spectra, lensing imprints non-Gaussianity into the CMB, and 
the CMB trispectrum encodes information about the 
deflection field power spectrum, or mapping of the photon positions, 
itself \cite{Hu:2001fa}.

\subsection{Theory \label{sec:thy}} 

Accurate codes exist for computing each of these power spectra, at least 
for the standard cosmology.  We utilize CMB\-easy, which already implements 
several useful extensions to further cosmological parameters, including 
neutrino masses and several classes of dark energy \cite{cmbeasy1,cmbeasy2}. 
We have crosschecked results (for constant dark energy equation of state) 
with another code, CAMB, to ensure accuracy.  
Numerical stability is crucial, because several groups of cosmological 
parameters are highly degenerate and the differences between the power 
spectra for different cosmologies can be small, so numerical noise can 
distort the results.  
We carry out 
parameter estimation through Fisher matrix analysis.  For the precision 
future data we consider, this should provide accurate constraints.  
We check for convergence of the final results for various step sizes of 
the cosmological model differencing. 

The set of parameters considered includes the standard ones of primordial 
perturbation amplitude $A_s$ and power law index $n$, optical depth 
$\tau$, physical baryon density $\Omega_b h^2$, cold dark matter density 
$\Omega_c h^2$ and dark energy density $\Omega_{de}$.  The Hubble 
constant is a derived parameter $h^2=(\Omega_b h^2+\Omega_c h^2)/ 
(1-\Omega_{de})$ under the assumption of spatial flatness.  The physical 
matter density is $\omega_m=\Omega_b h^2+\Omega_c h^2$.  Since 
neutrinos are known to have mass and this influences the lensing and 
other power spectra, we always include as a parameter the physical 
neutrino energy density $\Omega_\nu h^2$ or equivalently the sum of 
neutrino masses $\sum m_\nu=94(\Omega_\nu h^2)$ eV.  

Since no guarantee 
exists that dark energy is a cosmological constant, and generically 
other models have time variation of their equation of state, we consider 
two parameters, $w_0$ and $w_a$, to describe the dark energy equation of 
state, $w(a)=w_0+w_a(1-a)$.  Consideration of the physics behind dark 
energy led to this form \cite{linprl} and it has been shown to be 
accurate to $0.1\%$ in describing observables \cite{calib}.  The 
\lcdm\ model corresponds to fixing $w_0=-1$, $w_a=0$.  Given that the 
CMB has strong sensitivity to the early universe, we also consider 
another class of dark energy models, early dark energy, where the 
dark energy density is non-negligible around and before the recombination 
epoch.  These also have two parameters, the equation of state today 
$w_0$ and the constant high redshift early dark energy density $\Omega_e$ 
\cite{doranrob}.  For $z\lesssim2$ these look identical to the $w_0$-$w_a$ 
model where $w_a\approx5\Omega_e$ \cite{linrobb}, but have distinct and 
possibly significant effects at high redshift. 

Thus we simultaneously fit either seven or nine parameters. We use the 
following fiducial parameter values throughout the paper: $\{A_s,\, n,\, 
\tau,\, \Omega_b h^2,\, \Omega_c h^2,\, \Omega_{de},\, \sum m_\nu\} 
=\{2.41\times 10^{-9},\, 0.963,\, 0.084,\, 0.02255,\, 0.1176,\, 0.72,\, 
0.28 \,{\rm eV}\}$.

\subsection{Deflection Field \label{sec:defl}} 

The angular power spectrum of the CMB has been used to constrain 
cosmological parameters with unprecedented accuracy (see e.g.\ 
\cite{Komatsu:2008hk}), but its ability to inform us about the low redshift universe is limited by the so-called geometrical
degeneracy.
This arises because only angles are measured and, 
given some spectrum of primordial fluctuations, the power on each scale is nearly fixed for
constant $\sqrt{\omega_m}\,d_{\rm lss}$ (where $d_{\rm lss}$ is
the angular diameter distance to the CMB last scattering surface), which is degenerate under certain combinations
of late universe parameters.
(An exception to this arises on large angular scales because the integrated Sachs-Wolfe (ISW) effect \cite{Sachs:1967er} leaves
another signature of dark energy on large scales, however owing to cosmic variance this effect is of limited use.)
Also, the primordial CMB probes the baryon distribution at last scattering, which is smoothed on scales smaller than $\sim10'$ because
of Silk damping \cite{Silk:1967kq} in the last scattering surface, while massive neutrinos mostly impact matter agglomerations
on projected smaller scales. 

The geometrical degeneracy can be broken by adding for example Type Ia 
supernova (SN) 
distance information or constraints on the expansion rate to the CMB power spectrum constraint (see, e.g., \cite{Komatsu:2008hk}). The effect of neutrinos on small scale structure can be probed through galaxy clustering or the Lyman-$\alpha$ forest \cite{Tegmark:2006az,Seljak:2006bg}.  
Alternatively, or in addition, deflection of CMB photons on their way to us changes the statistics of the primordial pattern in a characteristic way that can be used to infer the lensing effect. What was originally a nearly Gaussian random field becomes non-Gaussian with the coherent correlation of patterns around large scale matter fluctuations. This type of non-Gaussianity, on a typical scale of approximately 2 degrees, is different from that used to study inflationary models \cite{Babich:2005en,Creminelli:2006gc}, in that its three point function vanishes on most scales (except for those large scales on which the unlensed CMB is correlated with the lenses through the ISW effect). 

Lensing is described by the displacement vector of CMB photons on the sky, 
$\mathbf{\alpha}(\theta)$, which is given as $\mathbf{\alpha}(\theta)=\frac{D_{\rm CMB}-D_{\rm lens}}{D_{\rm CMB}}\hat{\mathbf{\alpha}}(\theta)$ in terms of the deflection angle 
 \beq
 \hat{\mathbf{\alpha}} = \frac{4G}{c^2} \int d^2 x' \Sigma(\xv') \frac{\xv-\xv'}{|\xv-\xv'|^2}\, , \quad \Sigma (\xv) \equiv \int dD \,\rho(\xv,D)\,, 
\eeq 
where $D$ is the angular diameter distance.  
The vector $\xv$ describes the position in the lens plane, and the surface mass density (lensing can be imagined to good approximation as progressing through multiple, infinitely thin planes) is $\sum (\xv)$, a projection of the three-dimensional density field $\rho(\xv,D)$.  In the so-called Limber approximation, the lensing power spectrum $C_L^{\alpha \alpha}$
becomes a simple integral over the matter power spectrum at all redshifts weighted by angular diameter distance ratios. In this paper we refer to modes in the lensing power spectrum as $\Lv$ and $\lv$, to distinguish them from the CMB multipole $l$.  

Lensing also affects the angular power spectrum of the CMB  \cite{Seljak:1995ve}. The characteristic acoustic oscillation features are smeared out, as characteristically sized hot or cold spots are magnified or de-magnified by intervening lenses. The amount of over-smearing is scale dependent, encapsulating information about the shape of the matter power spectrum, which in turn is affected by dark energy properties and neutrino masses.  Because of the distance 
factors (the geometric kernel) and the growth factors, the matter power 
spectrum is best probed over the range $z\approx1-4$.   

The effect of lensing on the CMB power spectrum is calculated within \ceasy. In the presence of lensing,
the power spectrum variance is not of the trivial Gaussian random field form. The non-Gaussian covariance is
negligible in temperature and E-mode polarization because the relative effect of lensing on these is small,
however it is large for B-mode polarization, a factor of a few \cite{Smith:2006nk}.
The effects of marginalization when constraining individual parameters generally overwhelm the effect of
the excess covariance however \cite{Smith:2006nk, tristan}.  
We confirmed that the effect of non-Gaussian covariance on the parameter constraints in the next sections
is negligible by checking that the uncertainties change by less than 10\% 
(typically less than 1\%) if the sample variance in
the B-mode is increased by a factor of five.

The power spectrum over-smearing method provides a statistical estimate of lensing that is prone to sample variance because
the actual distribution of the lenses on the sky remains unknown. To reconstruct the lensing potential $\psi$ (the line
of sight projection of the gravitational
potential of which the deflection vector $\alpha$ is the gradient)
one needs to use
the non-Gaussian information imprinted into the CMB.  
Lensing conserves surface brightness, so the probability distribution function of temperatures remains unchanged.  Therefore 
the lowest order non-zero estimator of the lensing potential is quadratic.  
This has been investigated by \cite{Zaldarriaga:1998te,Guzik:2000ju} and the minimum variance estimator was given by \cite{Hu:2001tn}.
A quadratic estimator is generally of the form
\beq
\hat{\psi}(\Lv)=N(L) \int \frac{d^2 \lv}{2\pi^2}\theta(\lv) \theta'(\Lv-\lv) g(\lv,\Lv-\lv) \,,
\label{eq:quadest}
\eeq
where $\theta$ and $\theta'$ stand for temperature and/or polarization modes on the sky,
i.e.~$\theta, \theta' = T, E, B$.
The optimal weight $g$ and normalization $N$ can be found using the fact that the deflected position
can be written as a first order expansion of the displacement around the undeflected
position, $\theta^{\rm L}(\xv)=\theta^{\rm UL}(\xv+\alpha) = \theta^{\rm UL}(\xv) + \nabla^i\psi(\xv)\nabla_i\theta(\xv)$. 
For the TT estimator,
requiring an unbiased estimate and minimizing the variance leads to weighting of modes
\beq
g(\lv,\Lv-\lv) = \frac{(\Lv-\lv)\cdot \Lv C_{|\Lv-\lv|} + \lv\cdot \Lv C_l}{2 \tilde{C}_l^{\rm tot} \tilde{C}_{|\Lv-\lv|}^{\rm tot}} \,, 
\eeq
where $C_l$ ($\tilde{C}_l$) is the unlensed (lensed) temperature
power spectrum, following the convention
of\footnote{Note that other papers, for example \cite{Hu:2001tn, Hu:2001kj}, use the opposite notation to distinguish between lensed and
unlensed spectra.} \cite{LewChal06}.
Similar expressions follow for polarization.  The superscript ``tot'' 
originates from the fact that the lensed CMB and noise enter in the variance calculation.

With the definition in Eq.~(\ref{eq:quadest}) the noise of the lensing reconstruction equals the normalization which becomes
\beq
N(L)=\left[\int{\frac{d^2\lv}{2\pi^2}\left[(\Lv-\lv) \cdot \Lv C_{|\Lv-\lv|} + \lv \cdot \Lv C_l \right] g(\lv,\Lv - \lv)} \right]^{-1} \,. 
\label{eq:lensnoise}
\eeq 
Physically the noise is a combination of instrumental and intrinsic shape noise (see below). 

Note that this is only the best {\em quadratic} estimator.
Maximum likelihood methods can in principle be applied \cite{Hirata:2002jy,Hirata:2003ka}
but they have been shown to only give small improvements for temperature and polarization
experiments with the sensitivity levels assumed in this work, so we do not consider them here.
We also note that the approximation above leads to a bias in the quadratic estimator,
however for experiments considered in our paper, with angular resolutions larger
than 3' as well as noise levels down to a micro-Kelvin, these are only a few percent
and well understood (see \cite{Cooray:2002py,Hirata:2002jy}).

As is the case with the lensing of background galaxies, CMB lensing obtains most information
from the smallest scale resolved by any given experiment as these allow averaging over many
background features. Because shapes in the CMB temperature can be intrinsically elliptical,
averaging over many patterns becomes necessary to constrain relatively large lens features.
Since unlensed B-type polarization patterns should be absent on scales less than a degree or
so in concordance cosmology, quadratic estimators involving B, in particular the EB pair due
to its higher signal-to-noise, are intrinsically more useful than temperature
(as long as B can be imaged) and can be used to constrain lenses out to smaller scales.
Therefore experiments beyond {\sc Planck}, with the capability of imaging B-patterns,
allow for reconstruction of lenses out to smaller angular scales \cite{Hu:2001kj,Hirata:2003ka}. 

In the following sections we will compare the lensed power spectra 
method (i.e.\ the over-smearing of acoustic peaks) of inferring late universe parameter values to the optimal quadratic estimator (OQE) method.
In the latter case we will use constraints on the unlensed power spectra in conjunction with a forecasted constraint on the
lensing potential power
spectrum\footnote{Using the lensing potential power spectrum is equivalent to using
the deflection power spectrum. They are simply related by $C_L^{\alpha \alpha} = L^2 C_L^{\psi \psi}$ (in the
flat sky approximation applied here).}
$C_L^{\psi \psi}$ using Eq.~(\ref{eq:lensnoise}) so we do not count the lensing information twice.

\subsection{Experiments} 

In this paper we consider three different experiments, two of which are scheduled to begin observations in the near future, to forecast constraints on 
neutrinos and dark energy. The {\sc Planck} satellite will be launched in 
the second quarter of 2009 and will observe the full sky from the semi-stable Lagrange point L2. We take into account a foreground cut for galactic emission and assume a sky coverage of 0.75 to be useful for cosmological analysis. We have adopted the experimental specification values in \cite{Smith:2006nk}. 

Combining both large sky coverage and high sensitivity, we consider the futuristic {\sc CMBpol} concept of a satellite mission specialized on polarization with ultra-high sensitivity. We have used values from \cite{Zaldarriaga:2008ap}. Our assumed specifications are summarized in Table~\ref{table:exps}.  
We postpone further discussion of {\sc PolarBear}, an intermediate term and 
sensitivity experiment until Section~\ref{sec:inter}. 

\begin{table}[htbp]
\begin{center}
\begin{tabular*}{0.9\columnwidth}
{@{\extracolsep{\fill}} l c c c c c}
\hline
Experiment     & $\nu$ & $f_{\rm sky}$ & $\theta_{\rm FWHM}$ & $\Delta_T$ & $\Delta_P$ \\
\hline
Planck     & 100 GHz & 0.75 & $9.2'$ & 51 & - \\
		      & 142 GHz & 0.75 & $7.1'$ & 43 & 78 \\
		       & 217 GHz & 0.75 & $5.0'$ & 65 & 135 \\
		       \hline
PolarBear & 150 GHz &  0.025 & $4.0'$ & 3.5 & 5 \\
			& 220 GHz & 0.025 & $2.7'$  & 8.5 & 12 \\
			\hline
CMBpol & all freq.\ comb. & 0.75 & $3'$ & 1 & $\sqrt{2}$ \\
\hline
\end{tabular*}
\end{center}
\caption{Experimental specifications assumed in the forecasts in this paper, for the various frequency bands of {\sc Planck}, {\sc PolarBear}, and {\sc CMBpol}.
The temperature and polarization sensitivities $\Delta_T$, $\Delta_P$ 
are given in units of $\mu$K-arcmin.} 
\label{table:exps} 
\end{table}

From these experimental characteristics the full estimator covariance matrices for each multipole $l$
can be constructed (e.g.\ \cite{Zaldarriaga:1997ch}).
The (Gaussian) covariances between the power spectrum and cross correlation estimators are given by
\beqa 
\mathbb{C}(C_l^{XY}, C_l^{ZW}) &=&\frac{1}{(2l + 1) f_{\rm sky}} 
\bigl[\left(C_l^{XZ} + N_l^{XZ}\right) \times \nonumber \\
\left(C_l^{YW}+N_l^{YW}\right) \!\!&+&\!\! \left(C_l^{XW}+N_l^{XW}\right) \left(C_l^{YZ} + N_l^{YZ}\right) \bigr]\,,
\eeqa 
where the noise power spectrum\footnote{When there are multiple frequency bands, the total noise power spectrum is given by
$N_{l, tot}^{-1} = \sum_i N^{-1}_{l, i}$, where the sum is over the individual bands and we have suppressed the superscripts.}
\beq
N_l^{XX} = \left(\frac{\Delta_X}{T_0}\right)^2 e^{l(l+1)\theta_{\rm FWHM}^2/(8\ln 2)} \,, 
\eeq
for $XX = TT, EE, BB$, and $N_l^{\psi \psi}$ is given by
Eq.~(\ref{eq:lensnoise}) and $N_l^{XY}=0$ when $X \neq Y$. Here $\Delta_T$ and $\Delta_E = \Delta_B = \Delta_P$ are the
temperature and polarization sensitivities, $\theta_{\rm FWHM}$ is the angular resolution, and $T_0$ is the temperature of
the CMB today.

\begin{figure*}[!htb]
\begin{minipage}[t]{0.49\textwidth}
\centering 
  \includegraphics*[width=\linewidth]{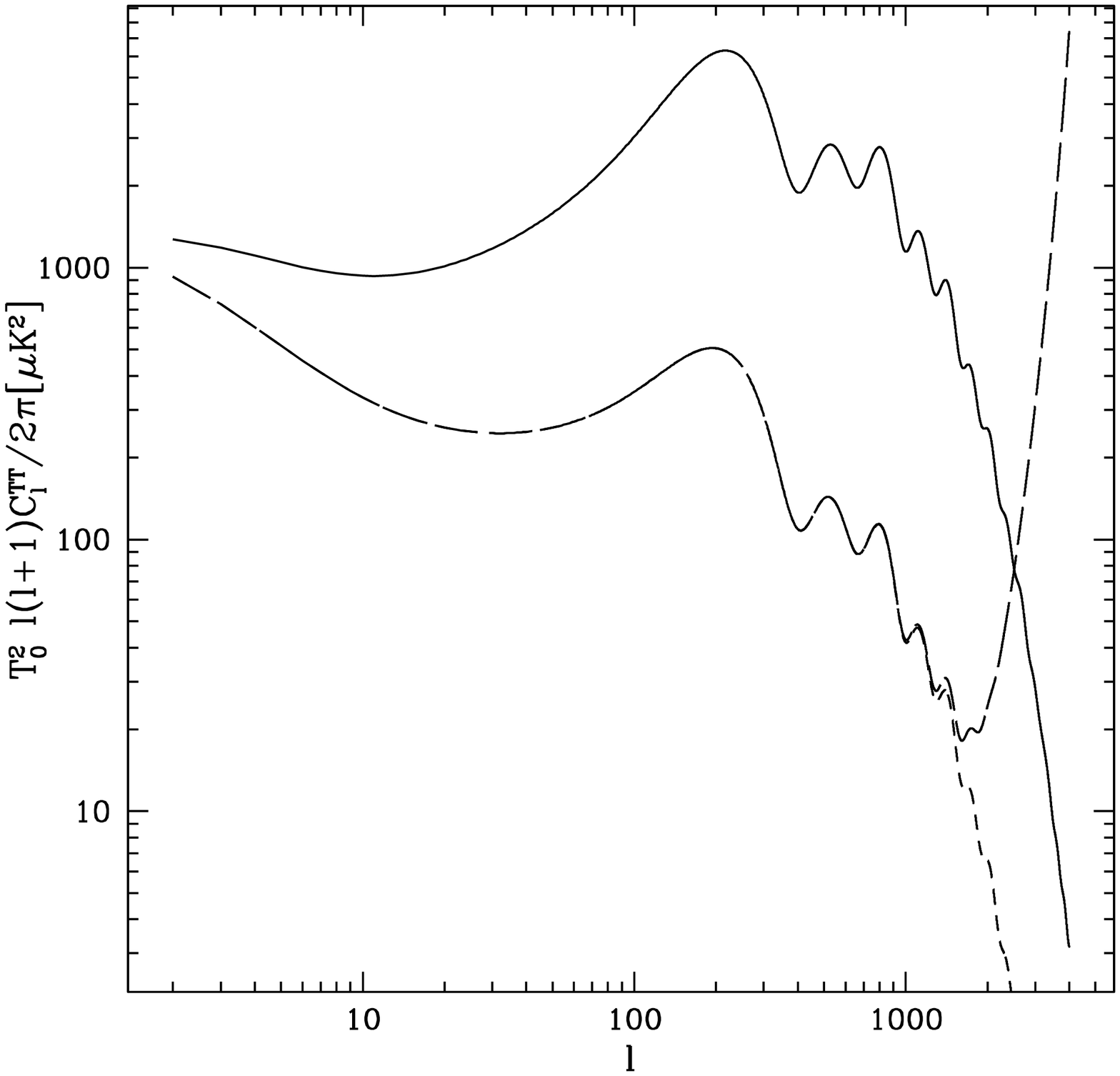}
  \includegraphics*[width=\linewidth]{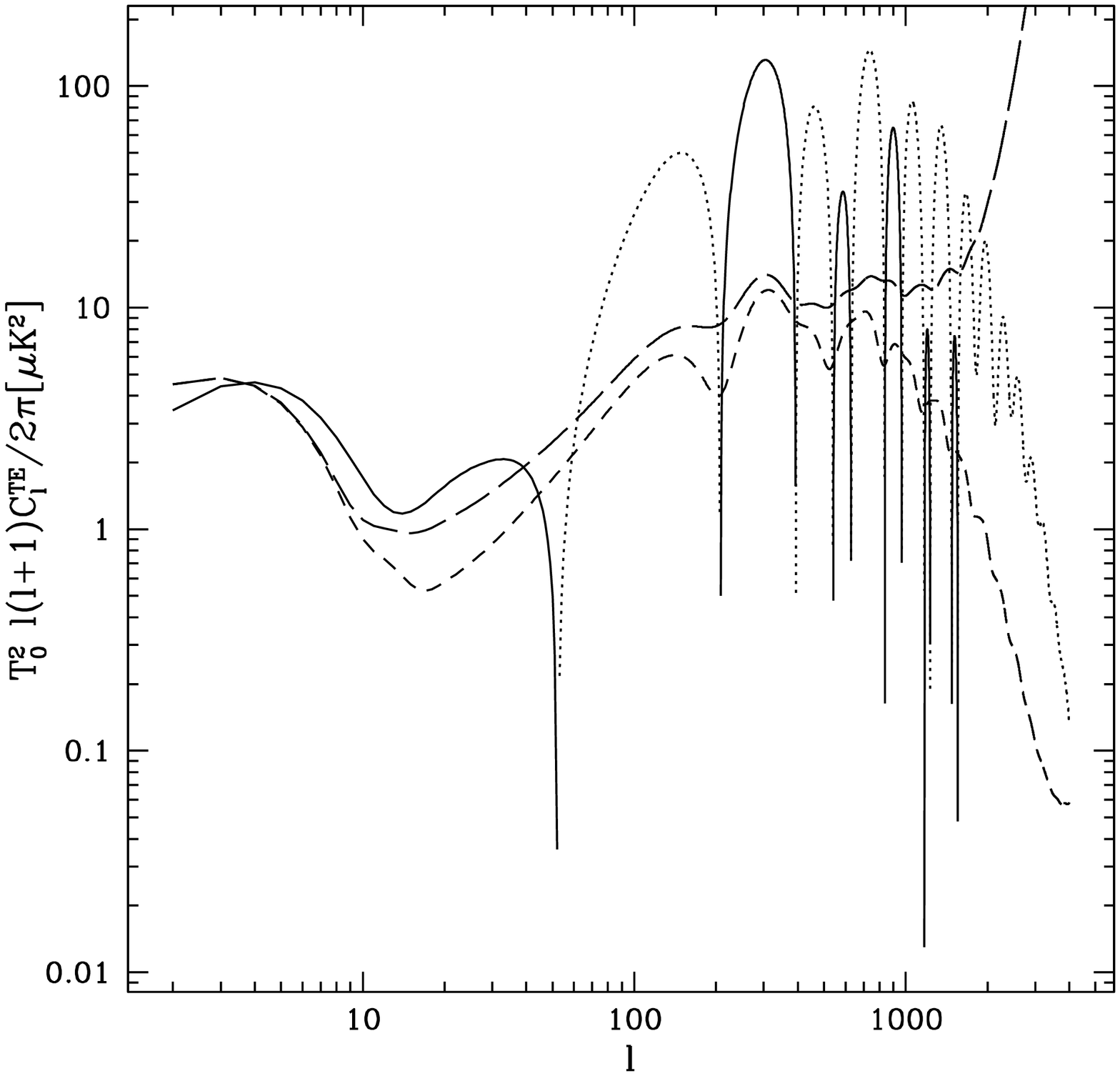}
\end{minipage} \hfill
\begin{minipage}[t]{0.49\textwidth}
\centering 
  \includegraphics*[width=\linewidth]{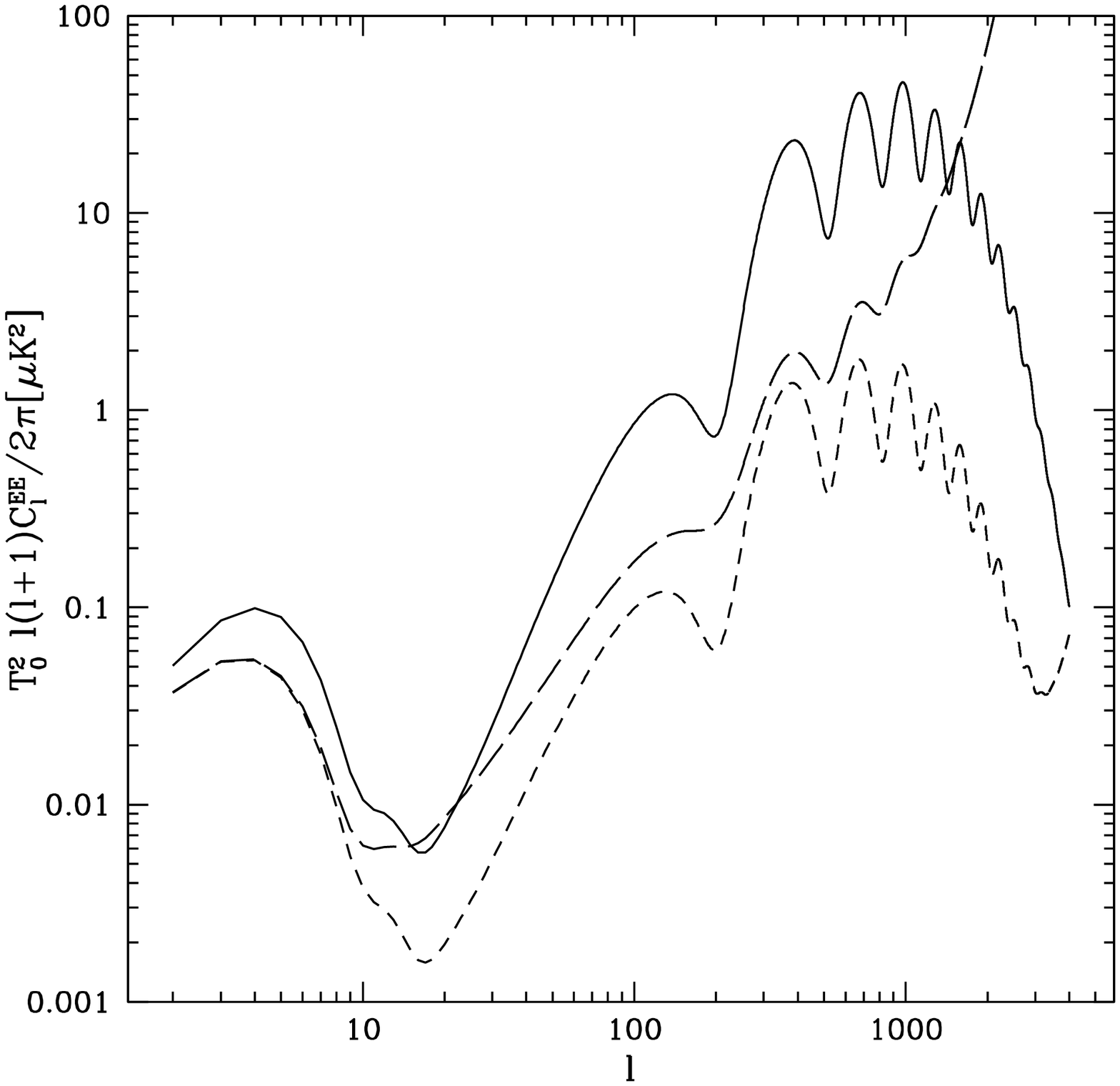}
  \includegraphics*[width=\linewidth]{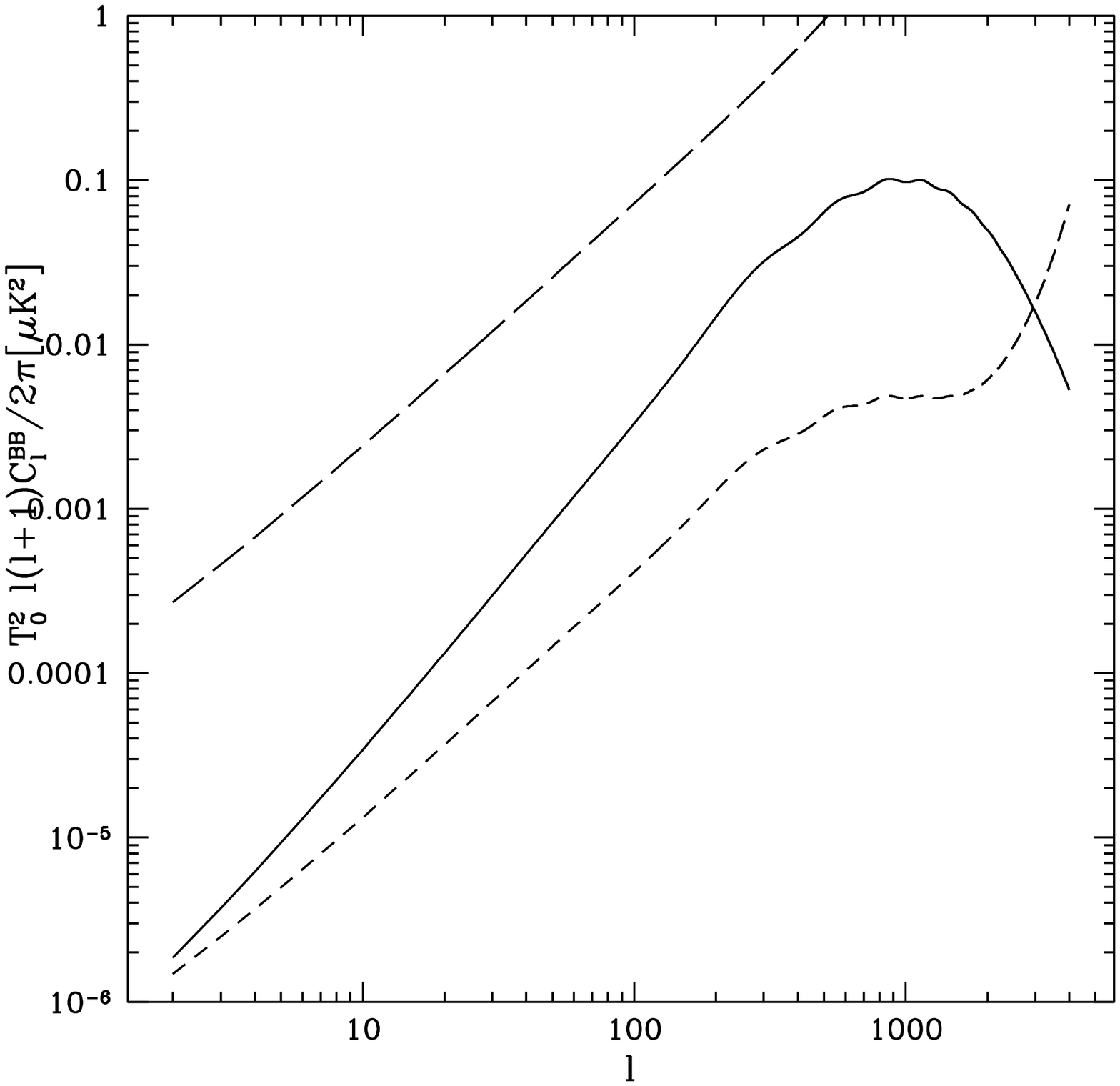}
\end{minipage} 
  \caption{Temperature and polarization power spectra
$T_0^2 l(l+1)C_l/2\pi [\mu K^2]$ vs.\ multipole $l$ for the \lcdm\ 
fiducial cosmology.  TT, EE, BB, and TE spectra (solid curves, dotted where 
negative) run clockwise from upper left.  
Dashed curves show the power spectrum errors, $T_0^2 l(l+1)\Delta C_l/2\pi [\mu K^2]$, for the Planck (long dash) and 
CMBpol (short dash) experiments.
}
\label{fig:spectra}
\end{figure*}

Figure~\ref{fig:spectra} shows that the noise of the experiments 
considered here is so low compared to the signal that they gather much 
of their information from scales beyond $l=2000$ in the temperature power 
spectrum ({\sc PolarBear} curves, not shown, would lie between {\sc Planck} 
and {\sc CMBpol} curves).  This is especially true for lensing, 
because 
the characteristic displacement of a CMB photon on its way from the last scattering surface to us is of order 2-3 arcminutes, and the smallest scale resolved by a given experiment contains most of the lensing information. 
However on scales $l\gtrsim 2000$ in the temperature power spectrum secondary anisotropies that are larger in magnitude than lensing, such as the Sunyaev-Zel'dovich (SZ) effects \cite{Sunyaev:1980nv} and radio as well as infrared point sources will make extraction of lensing information challenging. This is true as well for the optimal quadratic estimator, which might get confused by the extra non-Gaussianity carried by these foregrounds. In addition to the limitation due to the instrumental noise level and angular resolution, we therefore 
also quote our results with high-$l$ cuts at different scales, to show 
how these foregrounds affect parameter constraints. We note that while point sources and the SZ are expected to be significantly dimmer in polarization than in temperature \cite{Sazonov:1999zp,Hu:1999vq}, there the cutoff at high $l$ 
does not lead to as much loss in information as the polarization signal-to-noise ratio is small on angular scales beyond $l=2000$. 

Finally, the Fisher matrix is given by the expectation value of the second derivative of the logarithm
of the likelihood function $\mathcal{L}(C_l|\theta_i)$.
Assuming Gaussianity of the likelihood it is of the form
\begin{equation}                                                                  
F_{ij}=\sum_l \sum_{\alpha, \beta} \frac{\partial C^\alpha_l}{\partial                          
\theta_i}\mathbb{C}^{-1}(C^\alpha_l,C^\beta_l)\frac{\partial C^\beta_l}{\partial \theta_j},                                                                  
\end{equation}
where $\alpha$ and $\beta$ run over the five observables: temperature,
E-mode polarization, T-E cross correlation, B-mode polarization, and lensing potential power spectrum (where the OQE is used),
and $i,j$ run over the
cosmological parameters.  The covariance matrix between parameters is 
given by the inverse of the Fisher matrix. 

\section{Neutrino Mass Constraints in $\Lambda$CDM \label{sec:lcdm}} 

We begin looking at cosmological constraints in the simplest model 
consistent with both cosmological and local observations: a cosmological 
constant universe with non-zero mass neutrinos.  Three types of data 
cuts are employed -- by classes of observations, experiments, and 
systematics.  

The classes of observations are 1) unlensed TT, TE, EE 
power spectra, 2) adding the effect of lensing to 1), 3) adding the 
BB power spectrum to 2), and 4) using 1) plus information on the 
lensing potential through the optimal quadratic estimator 
discussed in \S\ref{sec:defl}.  This allows understanding of the 
effects of lensing on just the temperature and E-mode spectra, the 
information in just the BB power spectrum caused by lensing, and 
methods for using the complete effects of lensing. 

On the experimental side, we consider Planck, slated for launch in 
mid-2009, and the far future CMBpol mission.  Discussion of the impact 
of intermediate scale ground-based missions is postponed until 
Section~\ref{sec:inter}.  Additionally we examine the influence of 
the level of systematics in terms of $\lmax$, such as induced through 
foregrounds external to the experiments. 

\begin{figure*}[!htb]
\begin{minipage}[t]{0.49\textwidth}
\centering
  \includegraphics*[width=\linewidth]{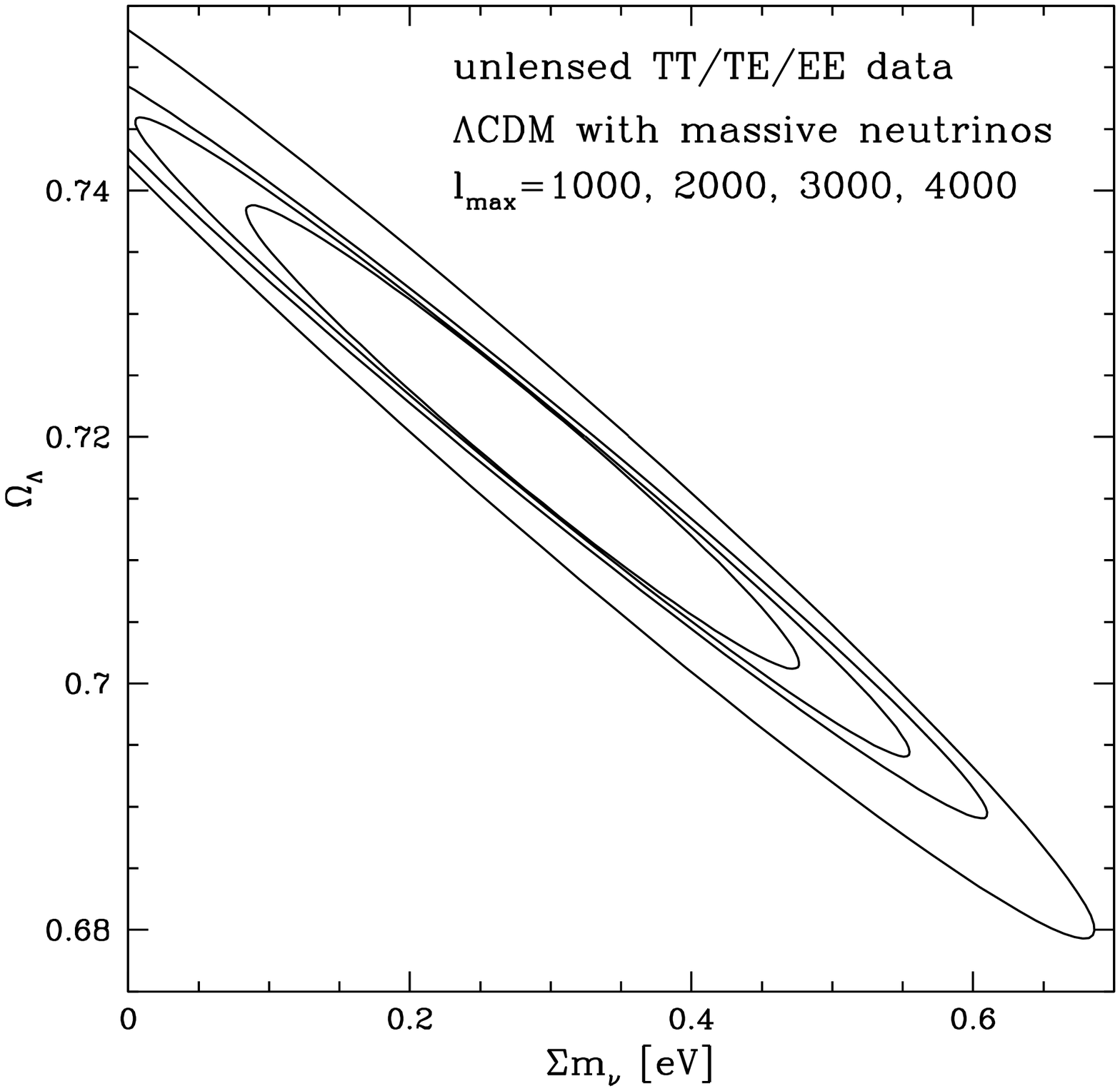}
  \includegraphics*[width=\linewidth]{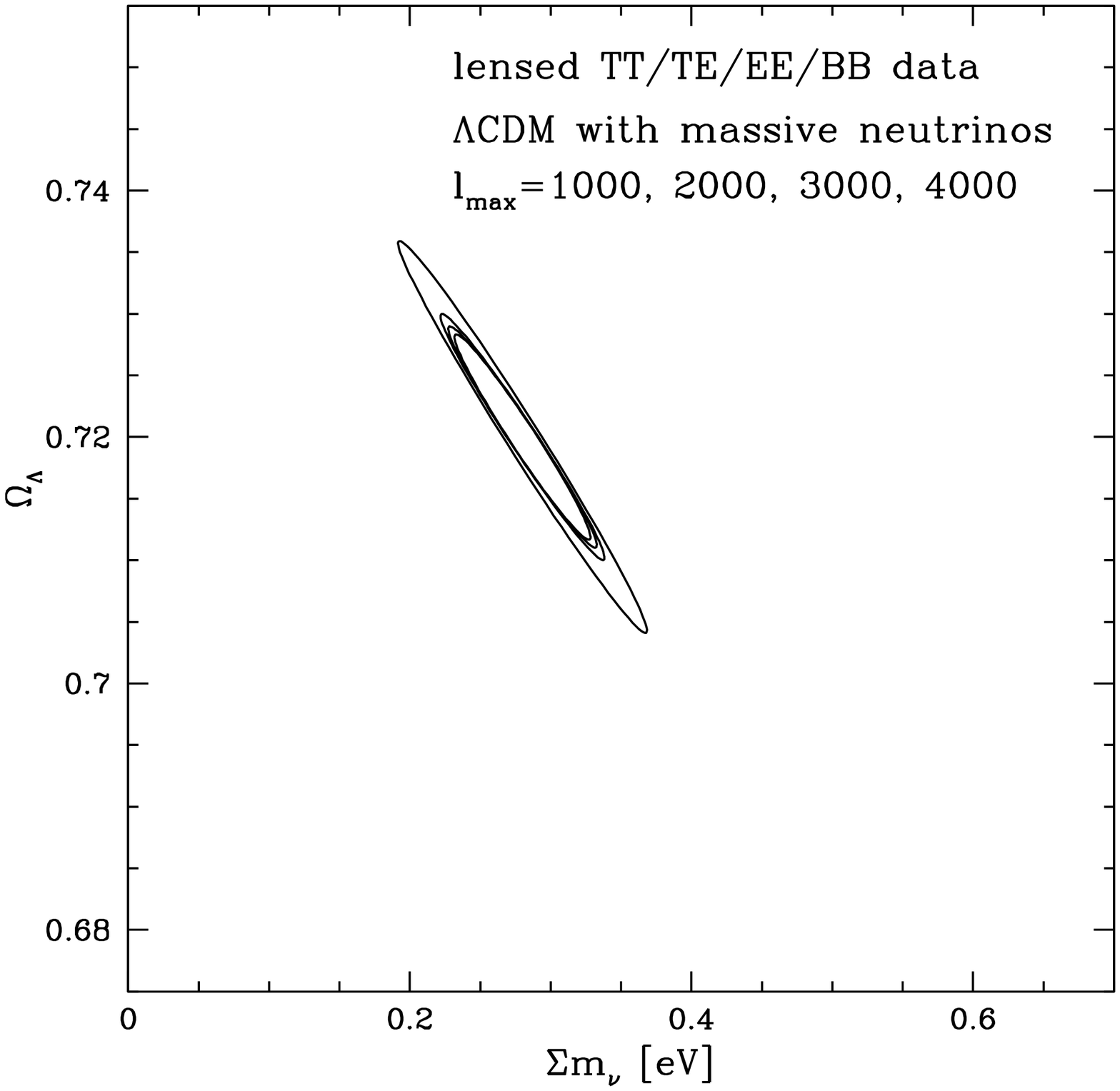}
\end{minipage} \hfill
\begin{minipage}[t]{0.49\textwidth}
\centering
  \includegraphics*[width=\linewidth]{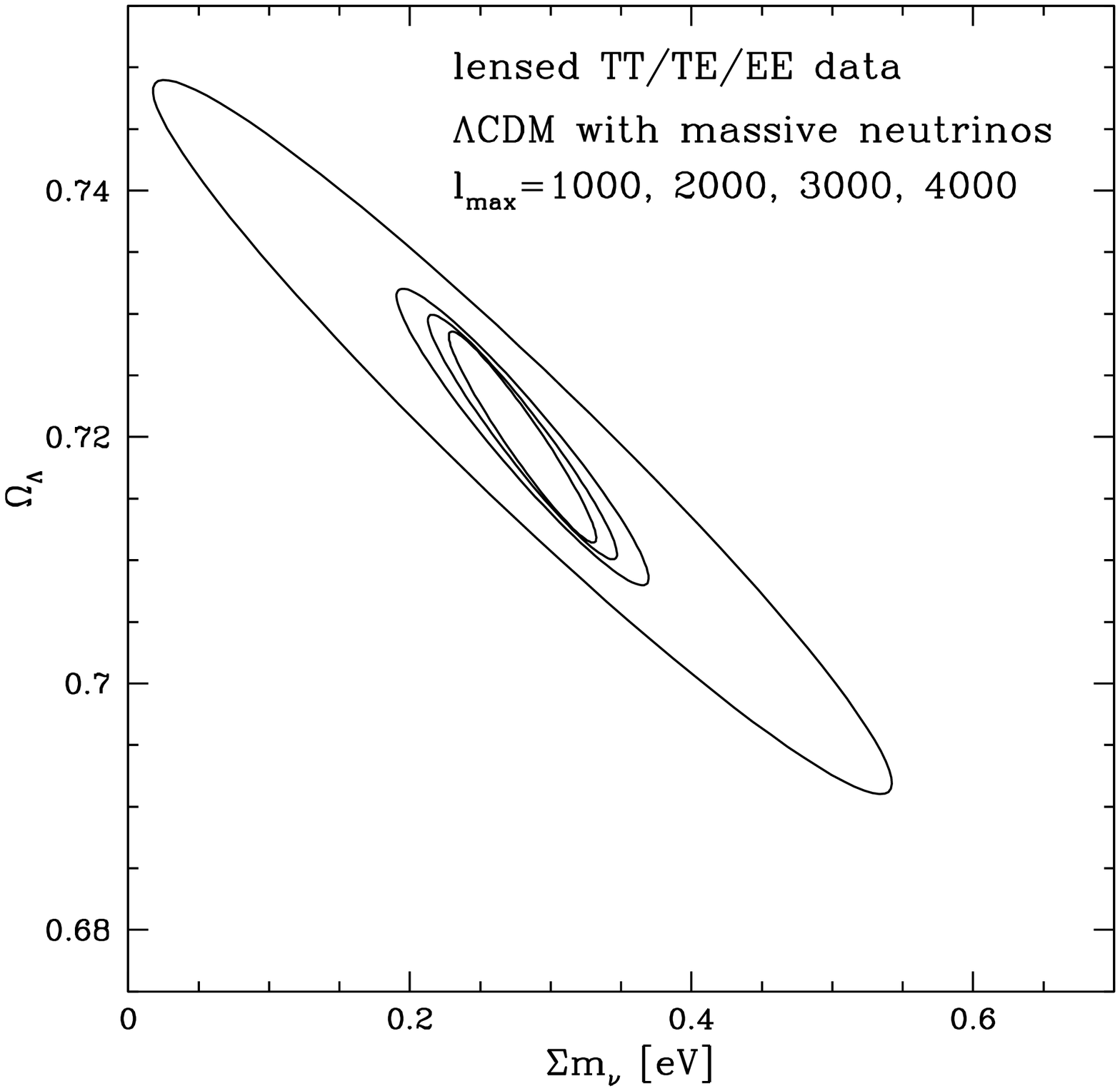}
  \includegraphics*[width=\linewidth]{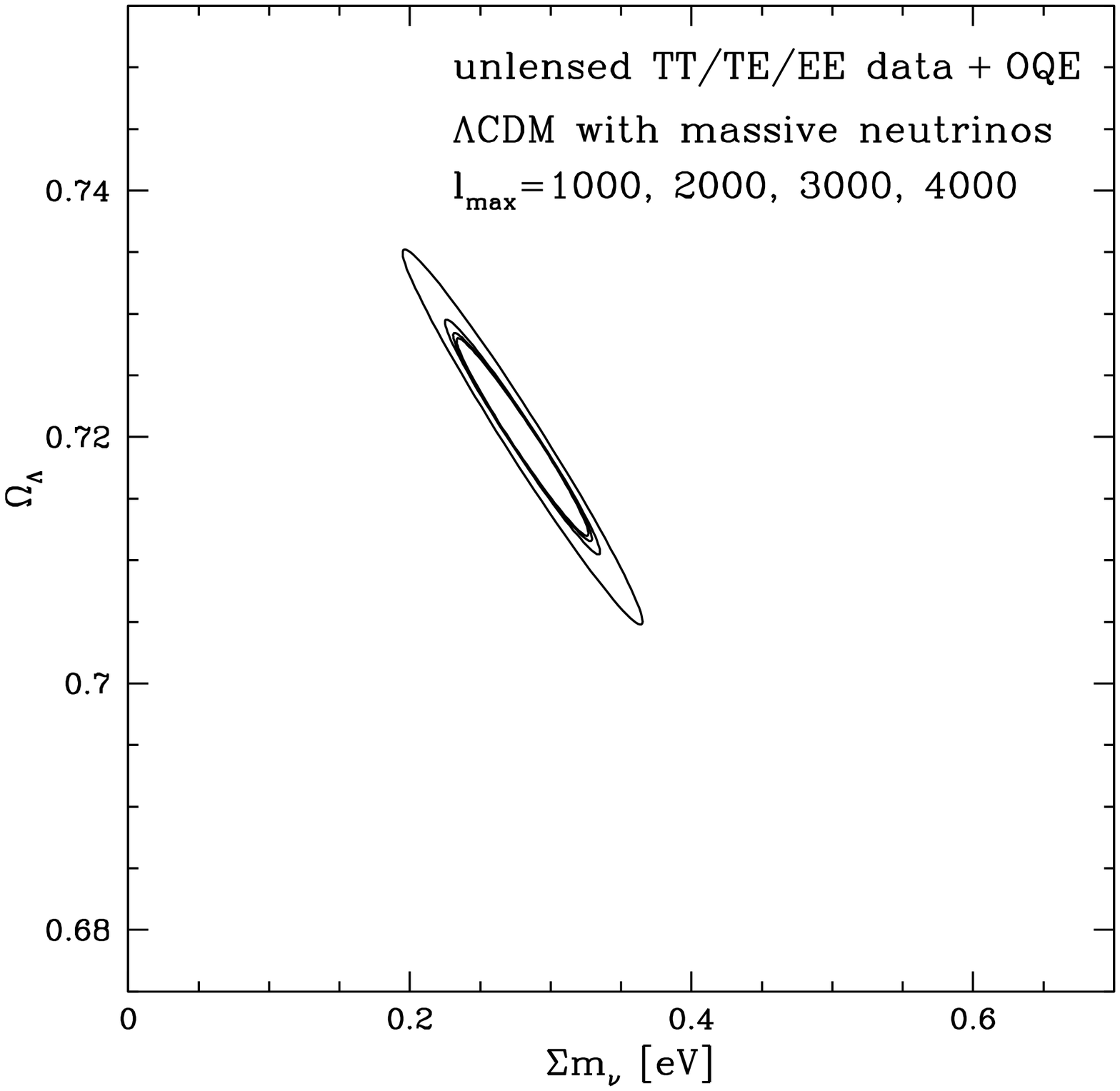}
\end{minipage}
  \caption{Cosmological constraints on the neutrino mass and dark 
energy density in the \lcdm\ fiducial cosmology from CMBpol.  Within 
each panel the contours correspond to systematic cuts at $\lmax=1000$, 
2000, 3000, 4000 from outer to inner.  The panels use different data cuts: 
no lensing (upper left), including lensing from T- and E-modes (upper right), 
including lensing from T-, E- and B-modes (lower left), and including lensing 
through the optimal quadratic estimator of the lensing potential (lower 
right).}
\label{fig:mnudata}
\end{figure*}

\begin{figure*}[!htb]
\begin{minipage}[t]{0.49\textwidth}
\centering
  \includegraphics*[width=\linewidth]{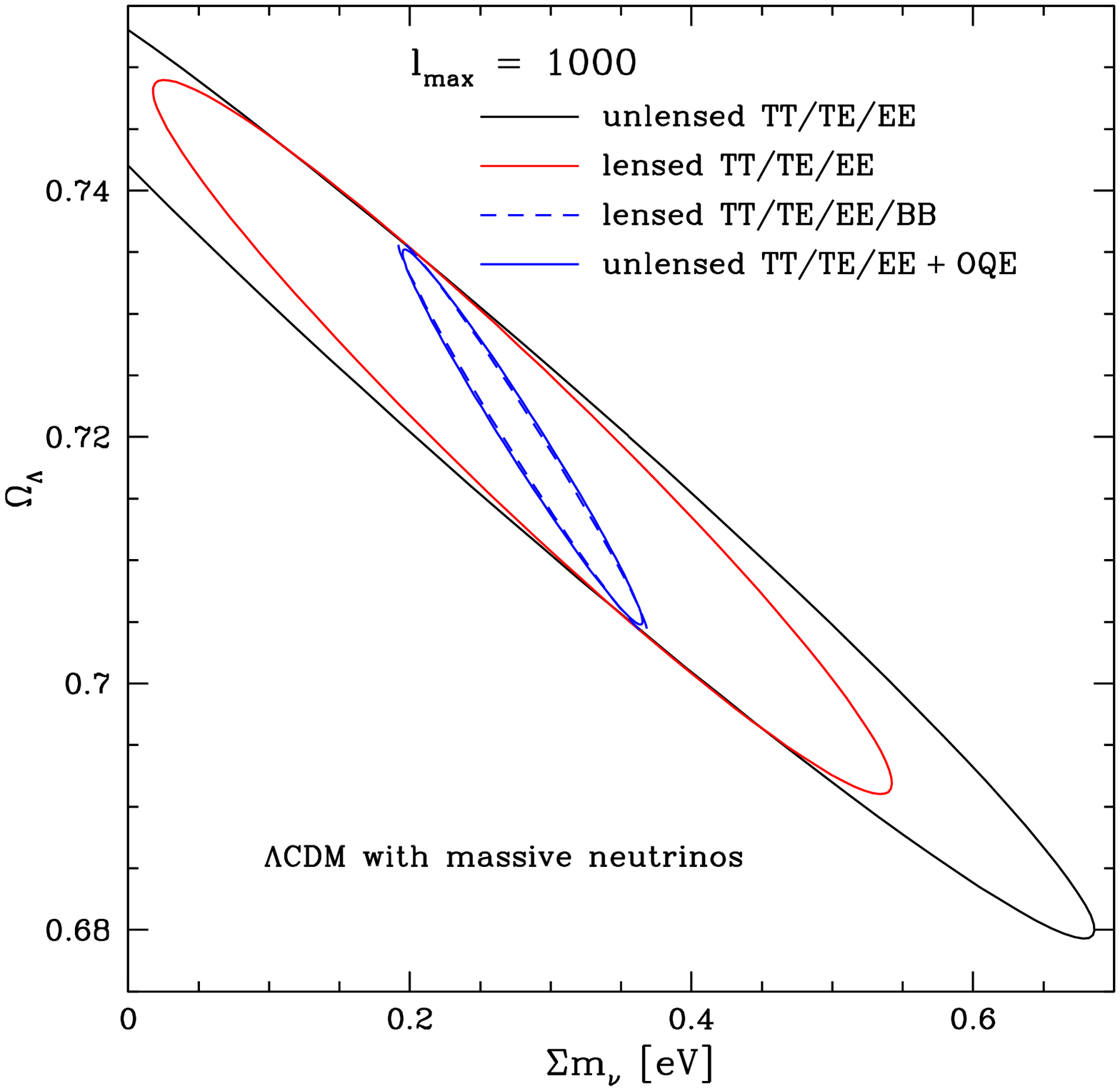}
  \includegraphics*[width=\linewidth]{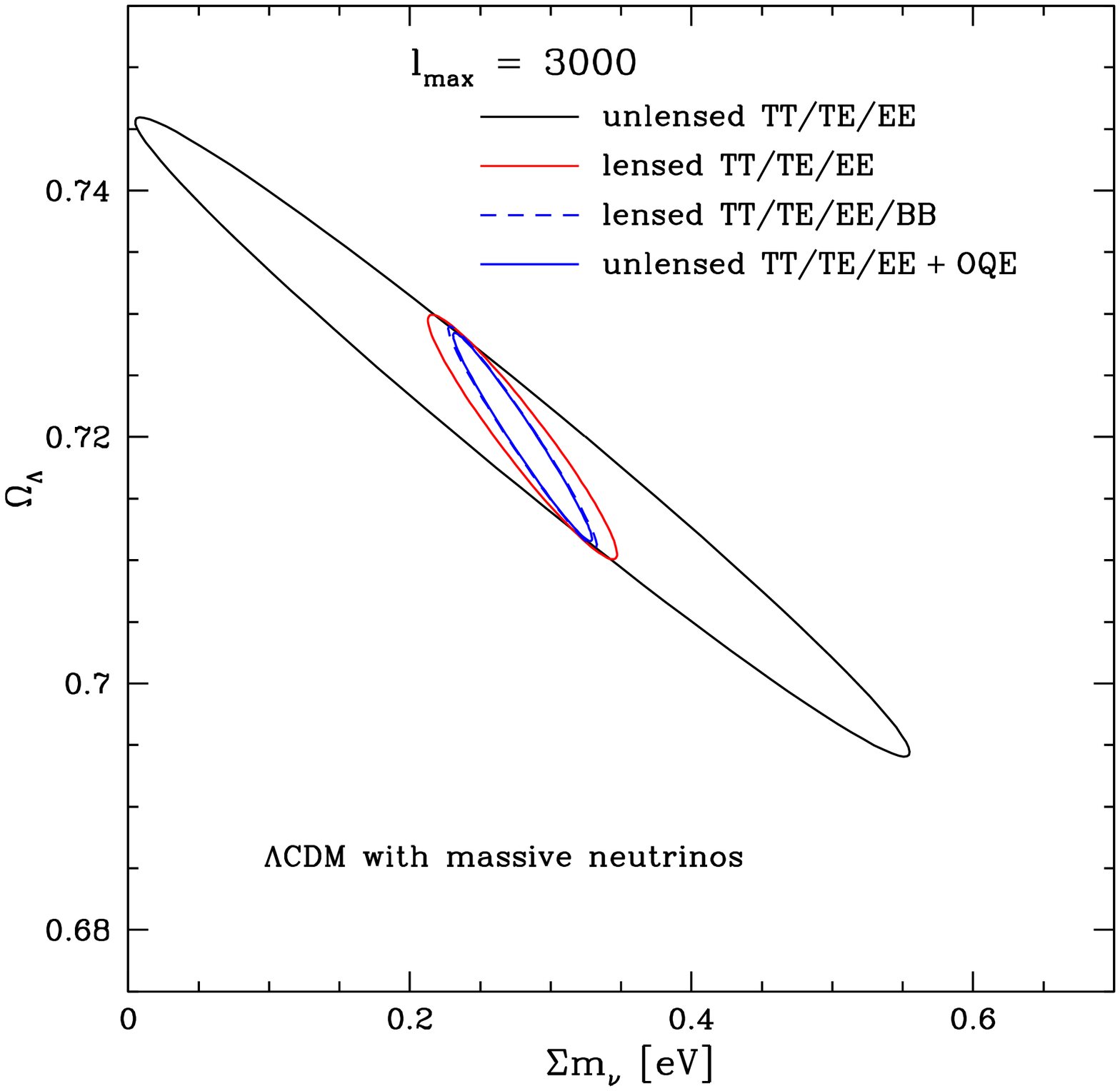}
\end{minipage} \hfill
\begin{minipage}[t]{0.49\textwidth}
\centering
  \includegraphics*[width=\linewidth]{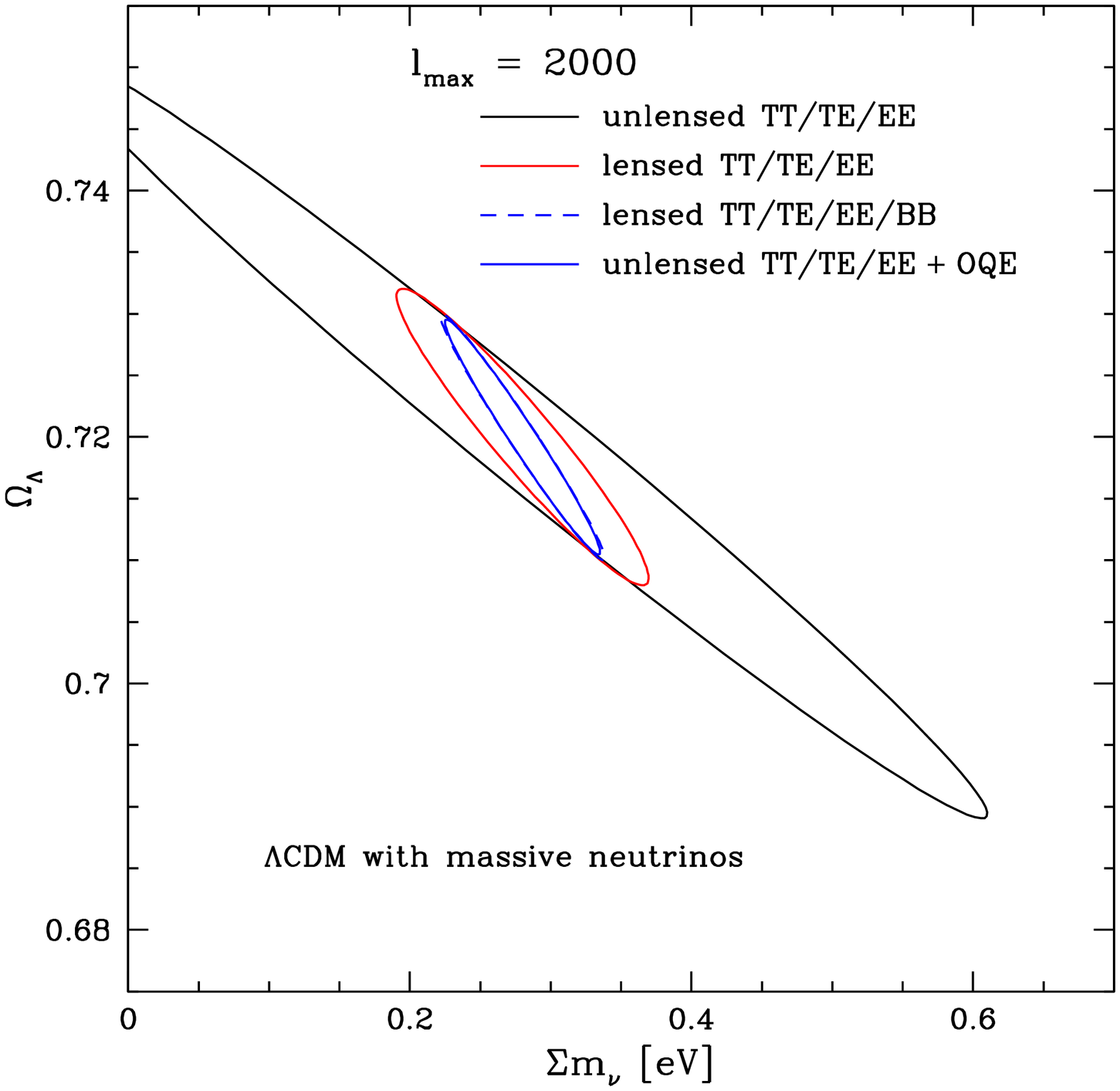}
  \includegraphics*[width=\linewidth]{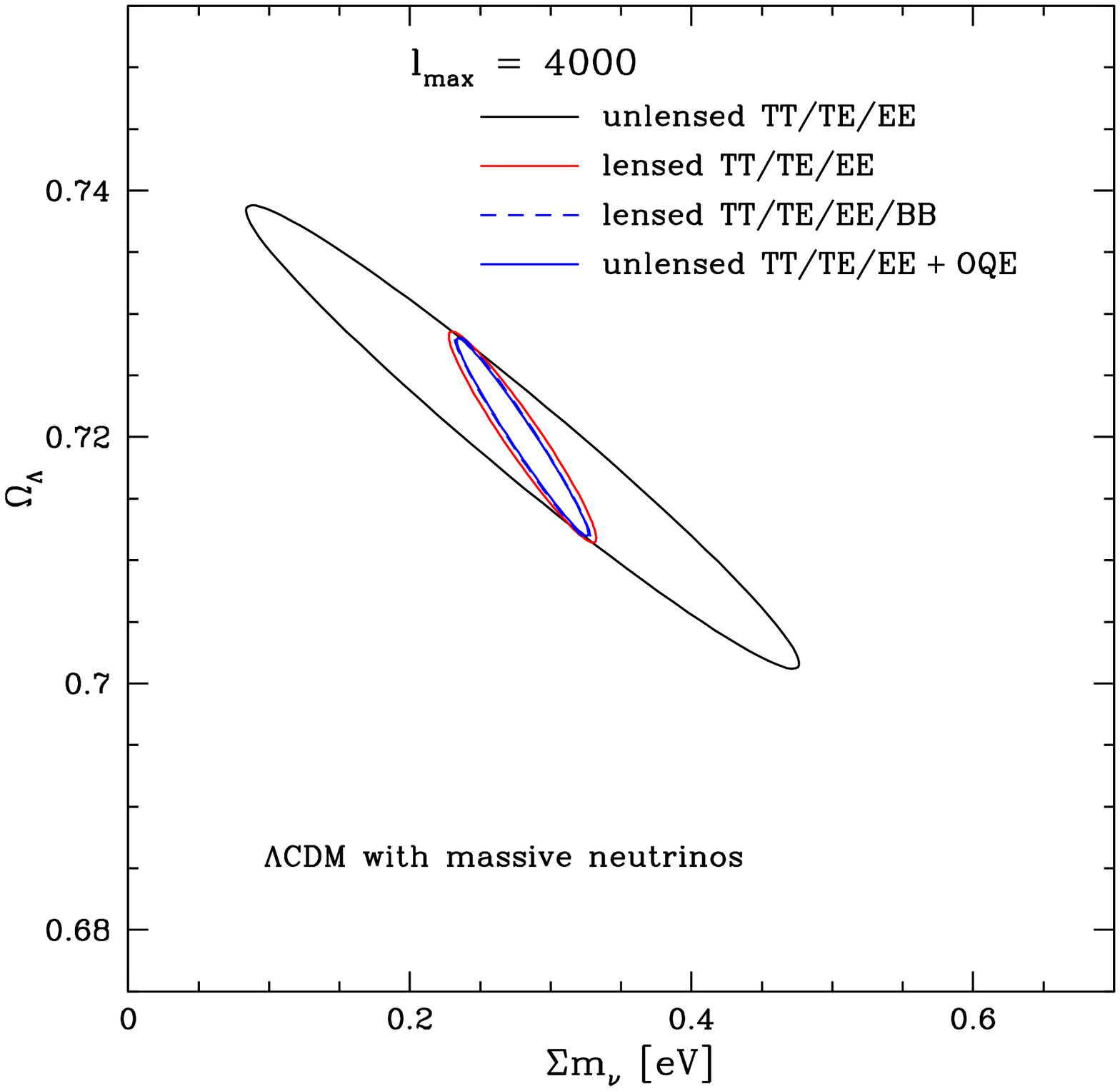}
\end{minipage} 
  \caption{As Fig.~\ref{fig:mnudata} but here within each panel the 
contours correspond to data set types, and the panels use different 
systematics levels: $\lmax=1000$ (upper left), 2000 (upper right), 
3000 (lower left), 4000 (lower right). 
}
\label{fig:mnulmax}
\end{figure*}

Figures~\ref{fig:mnudata} and \ref{fig:mnulmax} illustrate the constraints 
on neutrino mass and dark energy density (cosmological constant) for the 
different data set types and systematics levels.  All figures show 68\% 
confidence level contours; the fiducial 
model is \lcdm, with $\mnu=0.28$ eV.
Use of lensing information clearly adds substantial leverage, 
and measurement of B-modes or the lensing potential play an important 
role.  The two methods of including the full lensing information -- 
adding B-modes or adding the lensing potential -- are nearly equivalent 
(see \S\ref{sec:inter} for further discussion of this).

Considering the constraints for different systematics levels, 
we see that much of the lensing leverage 
is achieved by $\lmax\approx2000$. On smaller scales point sources and the SZ effects are expected to dominate over lensing and our limited ability to clean foregrounds through multifrequency observations will likely not allow lensing reconstruction on much smaller scales.  

\begin{figure}[!htb]
\begin{center}
 \includegraphics*[width=8.63cm]{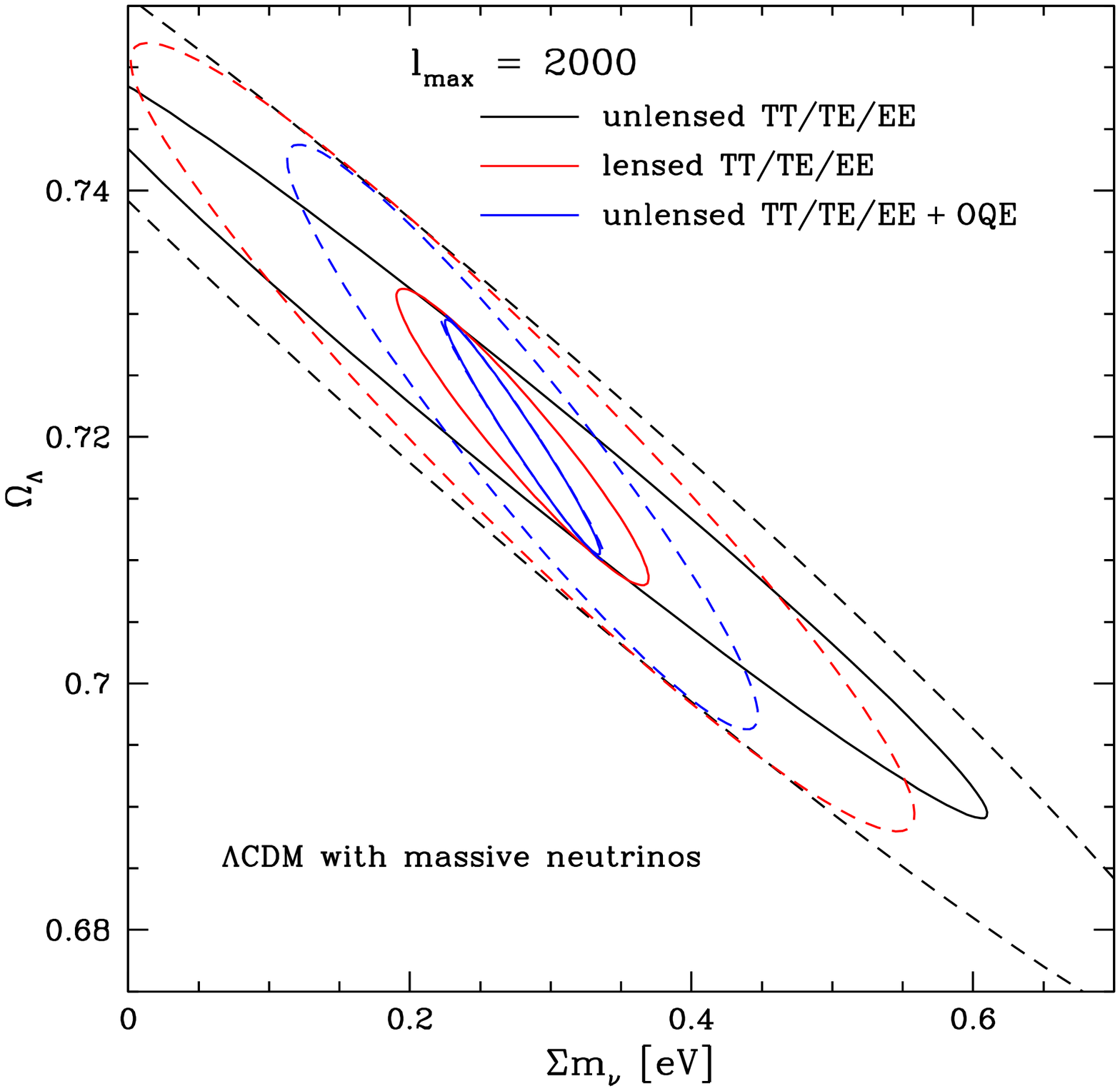}
\caption{Comparing the cosmological constraints on the neutrino mass 
and dark energy density in the \lcdm\ fiducial cosmology from 
Planck (dashed contours) vs.\ CMBpol (solid), taking $\lmax=2000$. 
}
\label{fig:mnuplanck}
\end{center}
\end{figure}

Finally, the dramatic improvement of CMBpol over Planck is clear 
in Fig.~\ref{fig:mnuplanck}.  Here we adopt as a standard systematics 
limit $\lmax=2000$ and show the confidence contours for each data set 
type for both experiments.   While lensing information does improve 
the Planck constraints, it runs into a wall due to the relatively 
high instrumental noise.  Furthermore,  Planck essentially cannot 
see B-mode lensing at all (see Fig.~\ref{fig:spectra}).  This is one of the motivations for 
intermediate experiments such as PolarBear. 

\section{Adding Dark Energy Dynamics \label{sec:wa}} 

The cosmic microwave background plays a crucial role in breaking 
the degeneracies of other probes in order to constrain the properties 
of dark energy.  However, unlensed CMB data itself has very little 
leverage on learning about dark energy, since the power spectra 
reflect mostly conditions in the high redshift universe or at best 
a single weighted average of dark energy influence through the 
distance to last scattering.  With the addition of lensed CMB data 
we can ask if this improves the leverage on dark energy; we emphasize 
that it is crucial to consider at least minimally realistic models 
that include dynamics in the EOS: taking the value of $w$ constant 
from the present to $z\approx1100$ is highly non-generic.  

It is also 
important to retain the inclusion of neutrino mass while making this 
investigation; both neutrino mass and dark energy influence the CMB 
in many of the same ways, e.g.\ suppressing structure and causing 
gravitational potentials to decay.  Ignoring neutrino mass could lead 
to overoptimistic constraints on dark energy.  In this section 
therefore we add $w_0$ and $w_a$ as fit parameters to the set considered 
in the previous section.  We explore the constraints under the same 
variety of data cuts as in that section. 

The geometric degeneracy due to the acoustic peaks feeling dark energy 
mostly through the integrated distance to last scattering remains 
strong, and no reasonable constraints can be placed on the dark energy 
EOS even with full use of the lensing information.  We therefore turn 
to the issue of complementarity: does the CMB data substantially help 
other probes of dark energy?  In particular we examine complementarity 
with luminosity distances measured by Type Ia supernovae, since the 
two probes are well known to strengthen each other \cite{fhlt,huhut}. 
We consider luminosity distances measured to $\sim1\%$ from $z=0-1.7$, 
including systematics, as could be provided by a supernova sample 
realized by a SNAP-type Joint Dark Energy Mission (JDEM)\cite{snap}.

\begin{figure*}[!htb]
\begin{minipage}[t]{0.49\textwidth}
\centering
  \includegraphics*[width=\linewidth]{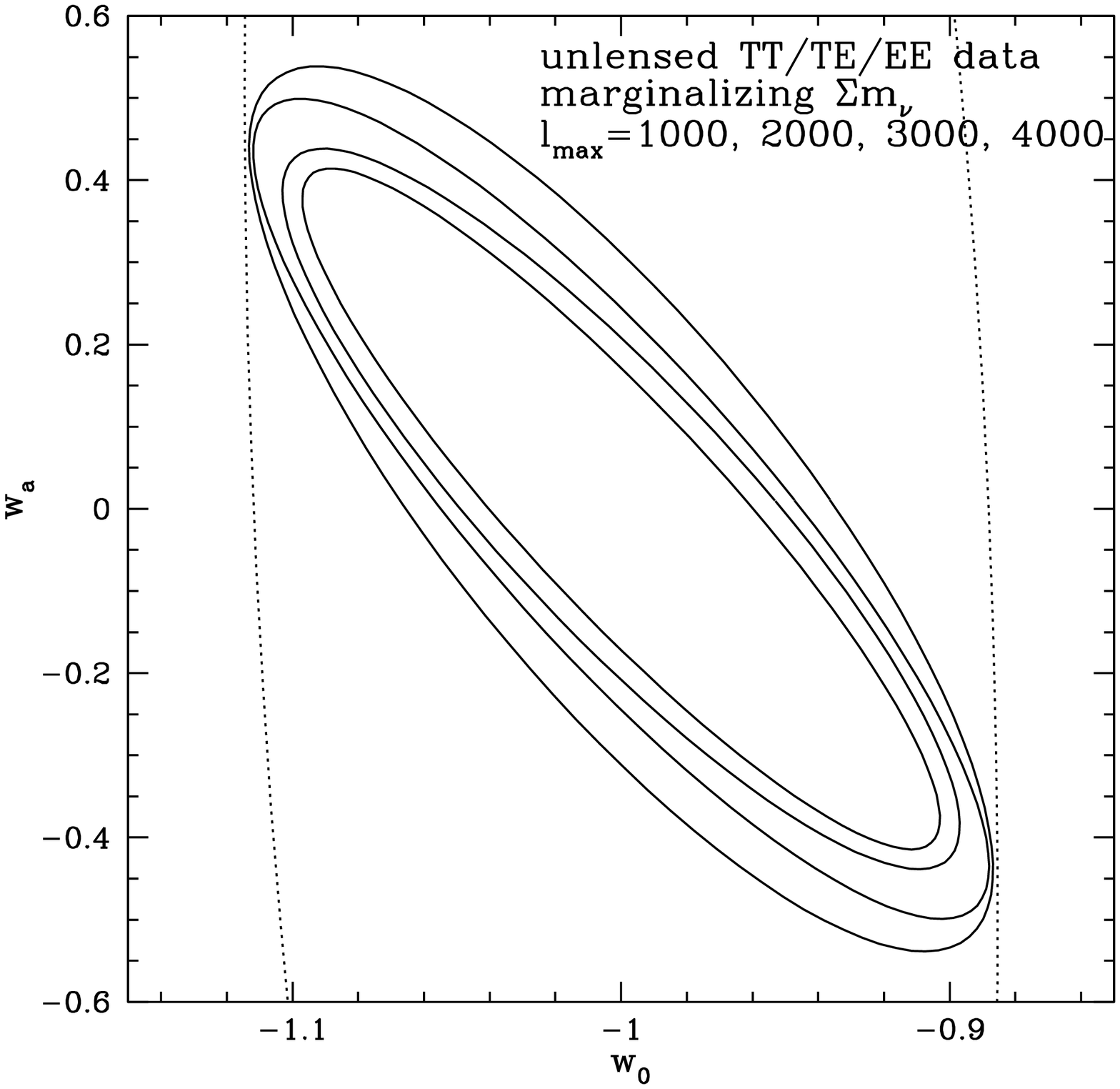}
\includegraphics*[width=\linewidth]{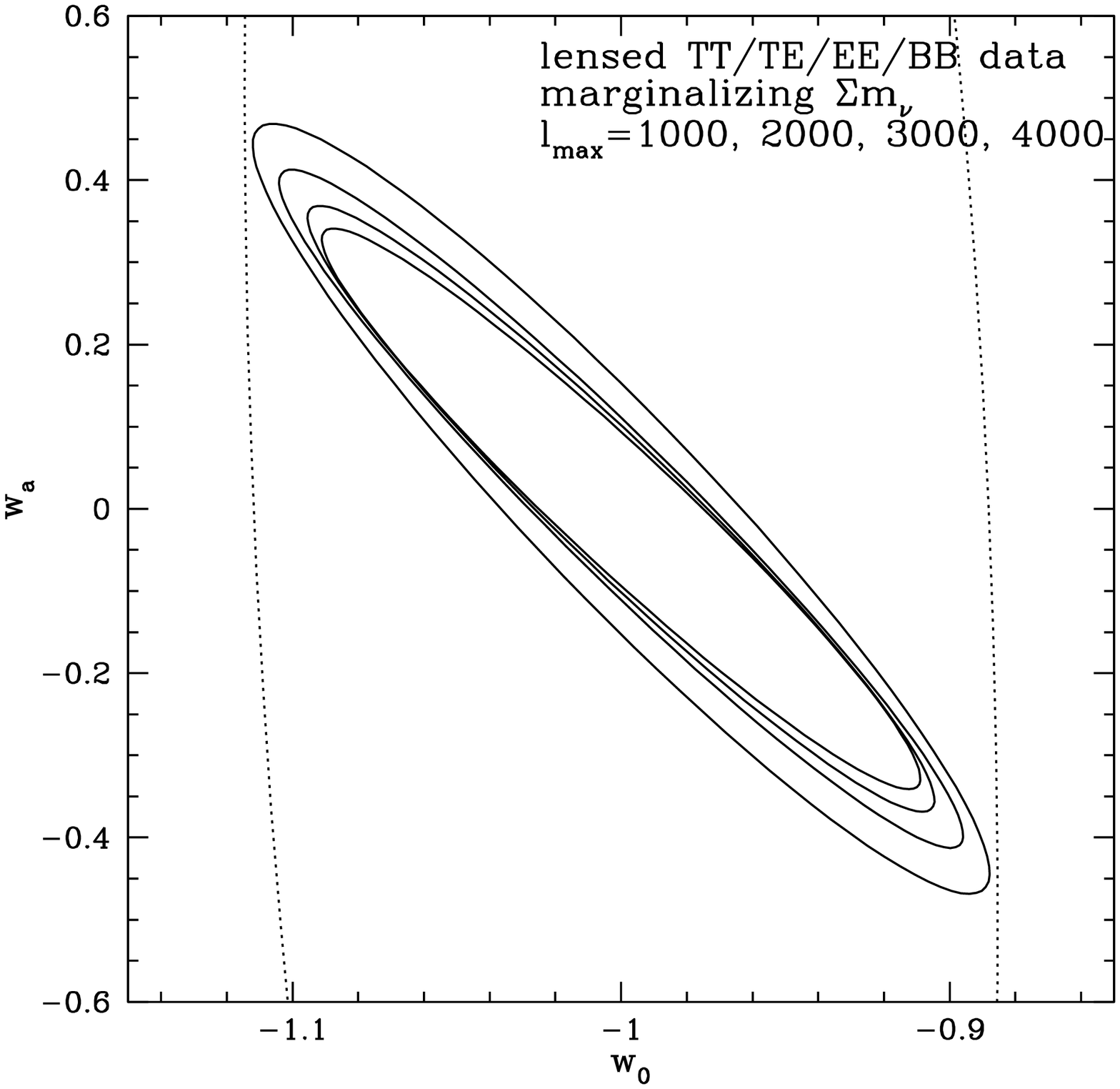}
\end{minipage} \hfill
\begin{minipage}[t]{0.49\textwidth}
\centering
  \includegraphics*[width=\linewidth]{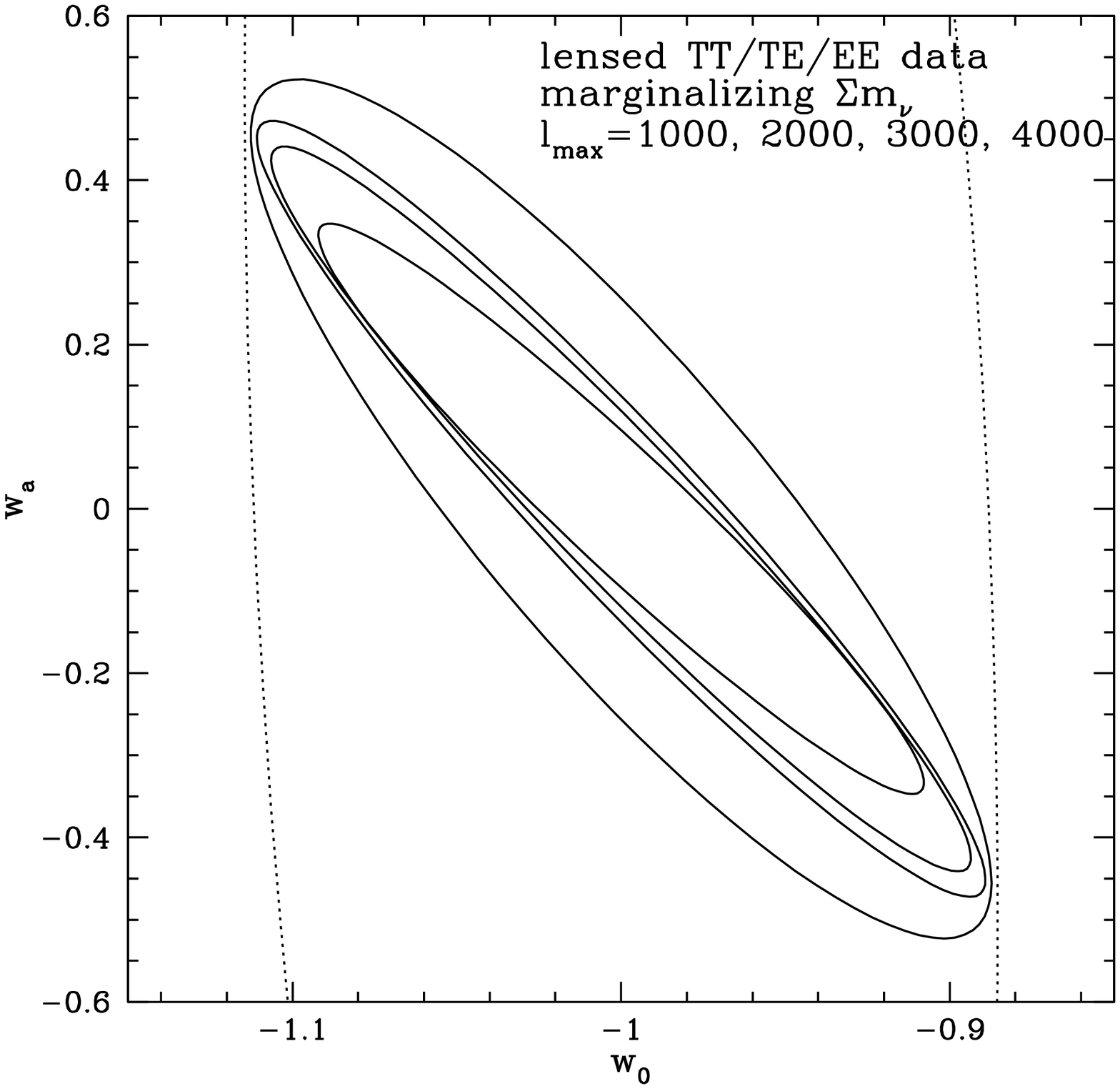}
  \includegraphics*[width=\linewidth]{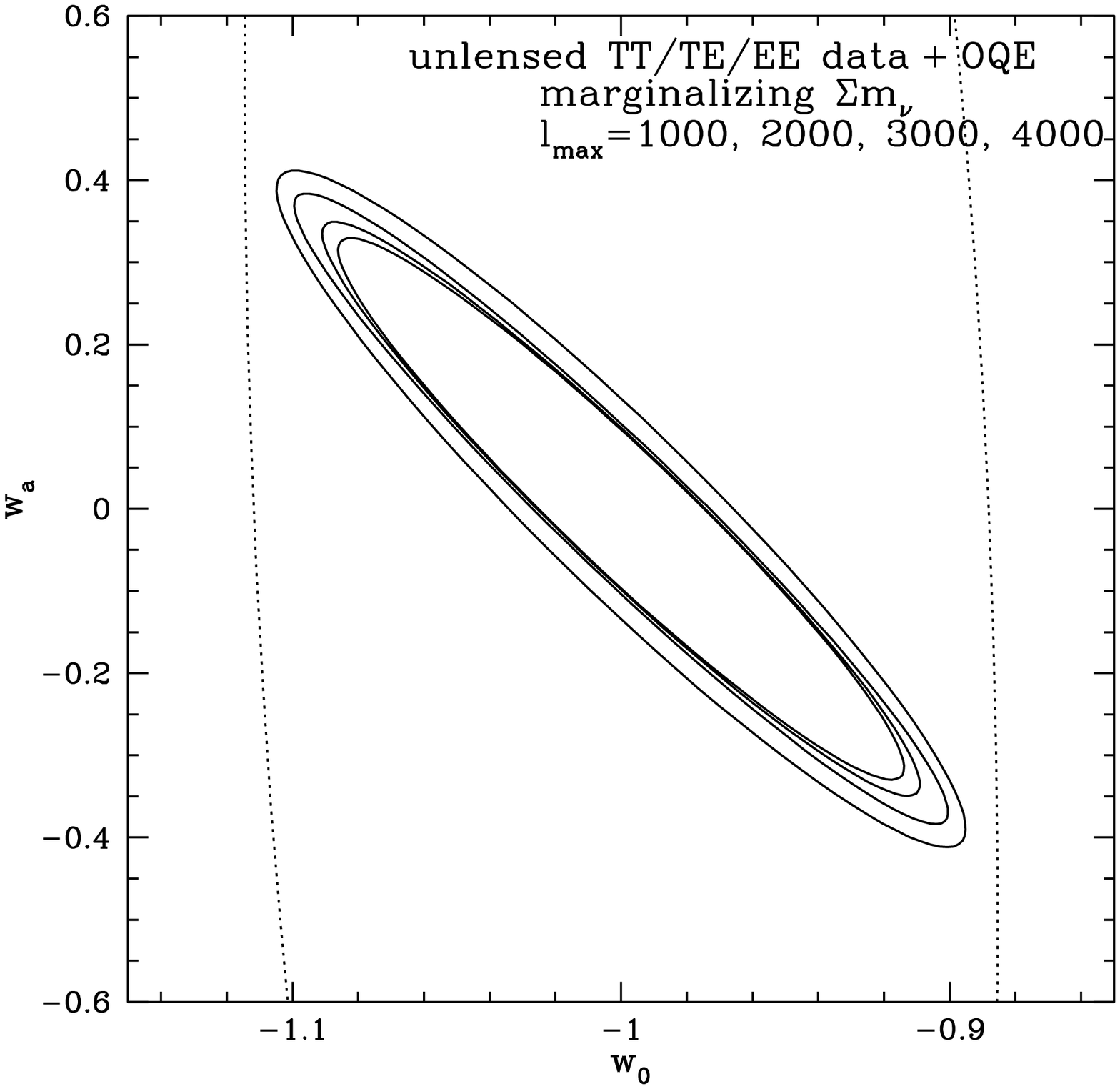}
\end{minipage}
  \caption{Cosmological constraints on the dark energy equation of 
state parameters $w_0$ and $w_a$ from CMBpol in combination with 
SNAP-quality supernova distances.  Within
each panel the contours correspond to systematic cuts at $\lmax=1000$,
2000, 3000, 4000 from outer to inner.  The panels use different data cuts:
no lensing (upper left), including lensing from T- and E-modes (upper right),
including lensing from T-, E- and B-modes (lower left), and including lensing
through the optimal quadratic estimator of the lensing potential (lower
right).  The dotted curve gives the constraints from supernovae alone. 
}
\label{fig:wadata}
\end{figure*}

Figure~\ref{fig:wadata} shows the constraints in the $w_0$-$w_a$ plane, 
marginalizing over the other seven parameters, for each data 
set type.  The first thing to 
notice is the clear improvement in measuring the time variation $w_a$ 
over the supernova sample alone due to even unlensed CMB data.   Adding 
lensed CMB data continues to tighten the constraints, in both $w_0$ and 
$w_a$, except for the worst systematics level $\lmax=1000$.  Full 
lensing information continues the improvement modestly on the limits, 
and somewhat narrows the contours.

\begin{figure*}[!htb]
\begin{minipage}[t]{0.49\textwidth}
\centering
  \includegraphics*[width=\linewidth]{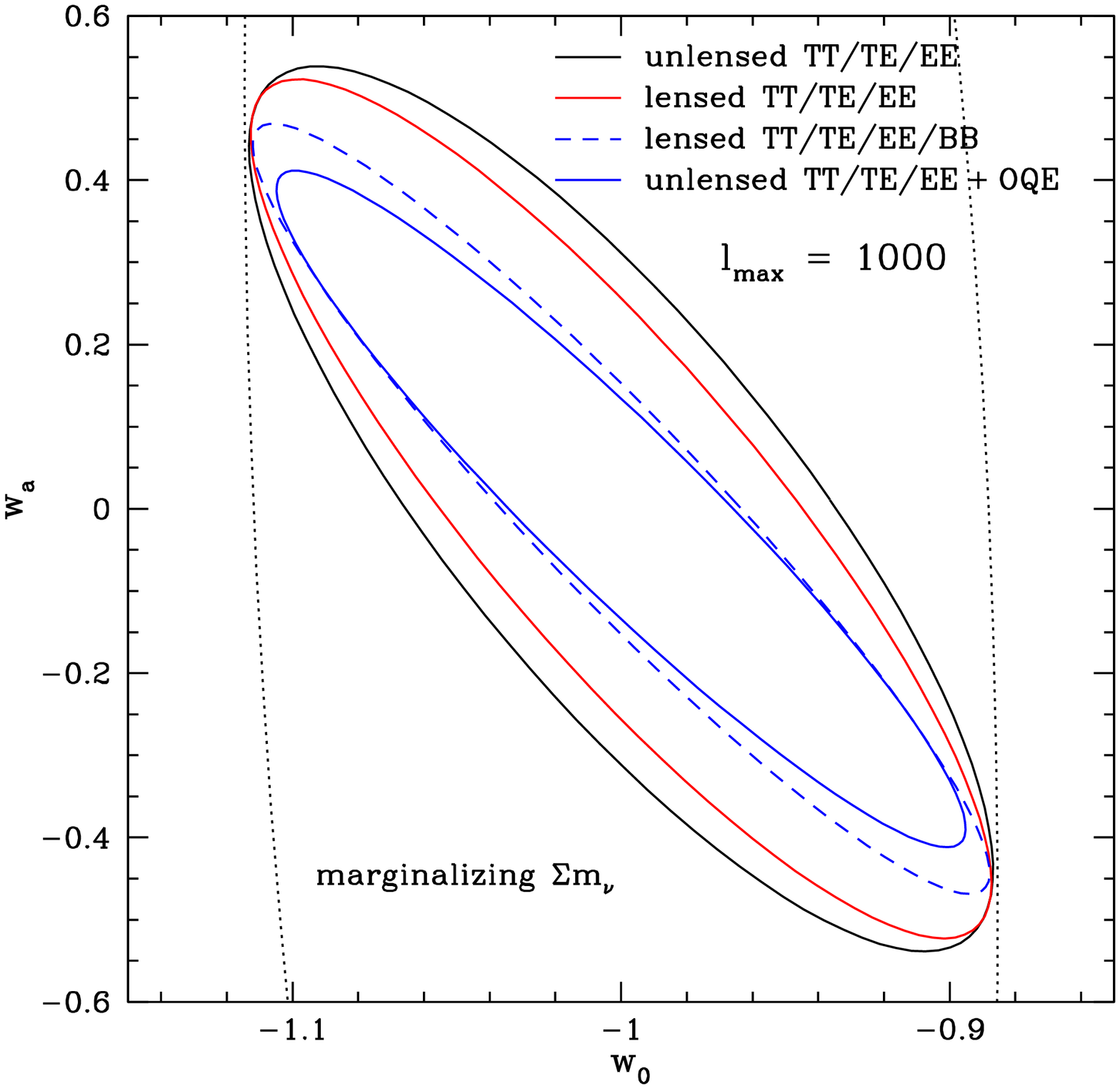}
\includegraphics*[width=\linewidth]{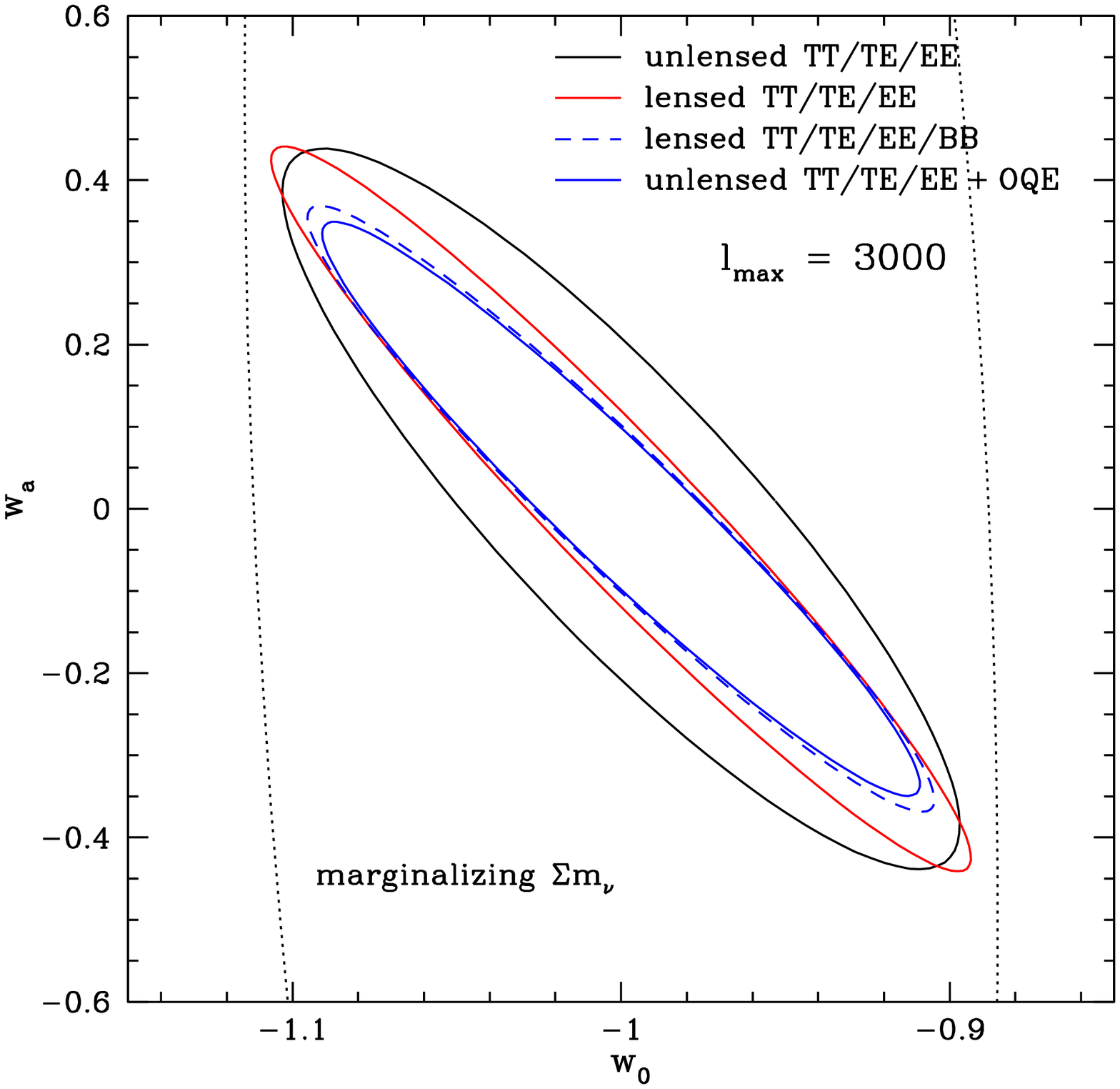}
\end{minipage} \hfill
\begin{minipage}[t]{0.49\textwidth}
\centering
  \includegraphics*[width=\linewidth]{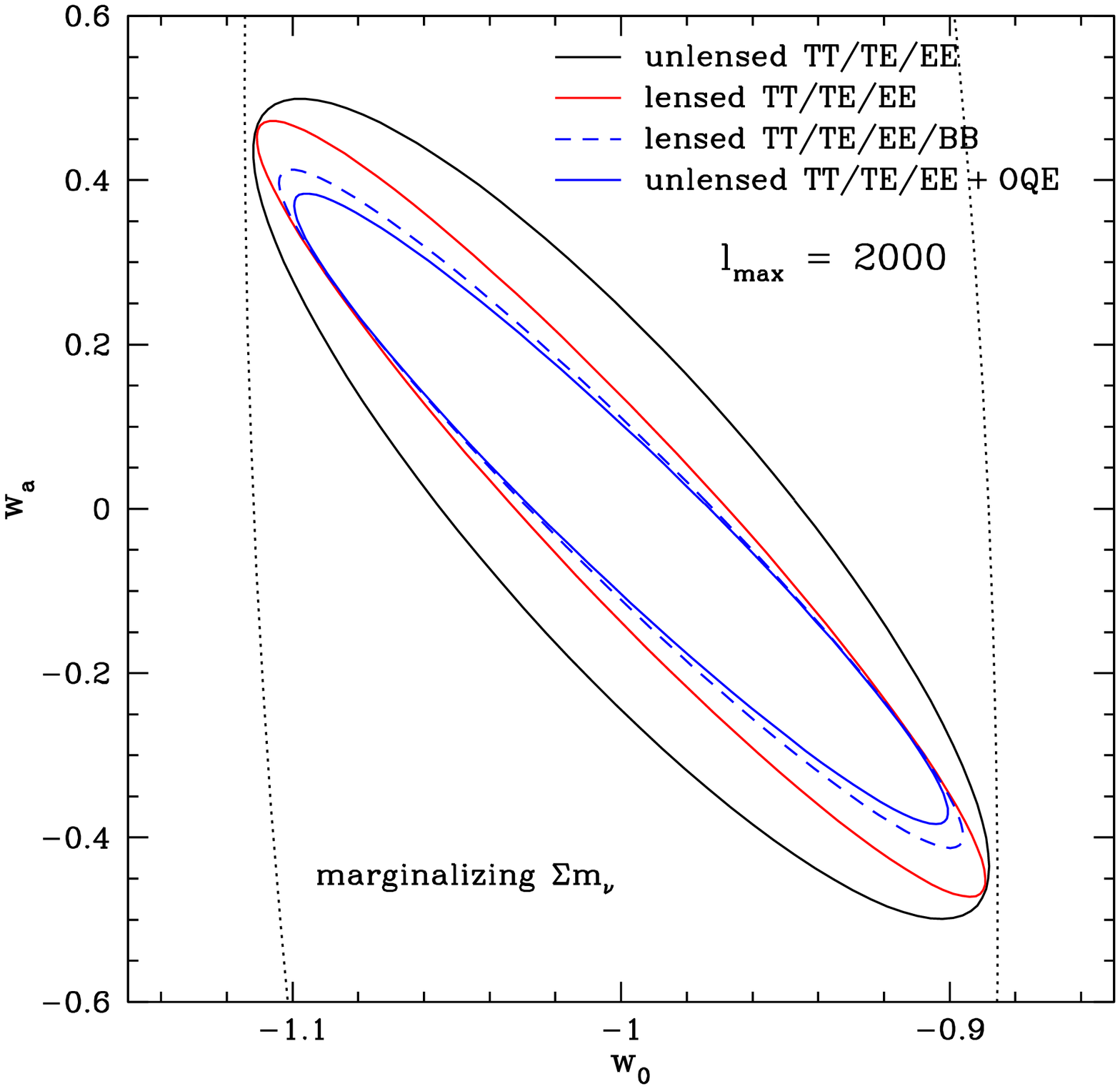}
  \includegraphics*[width=\linewidth]{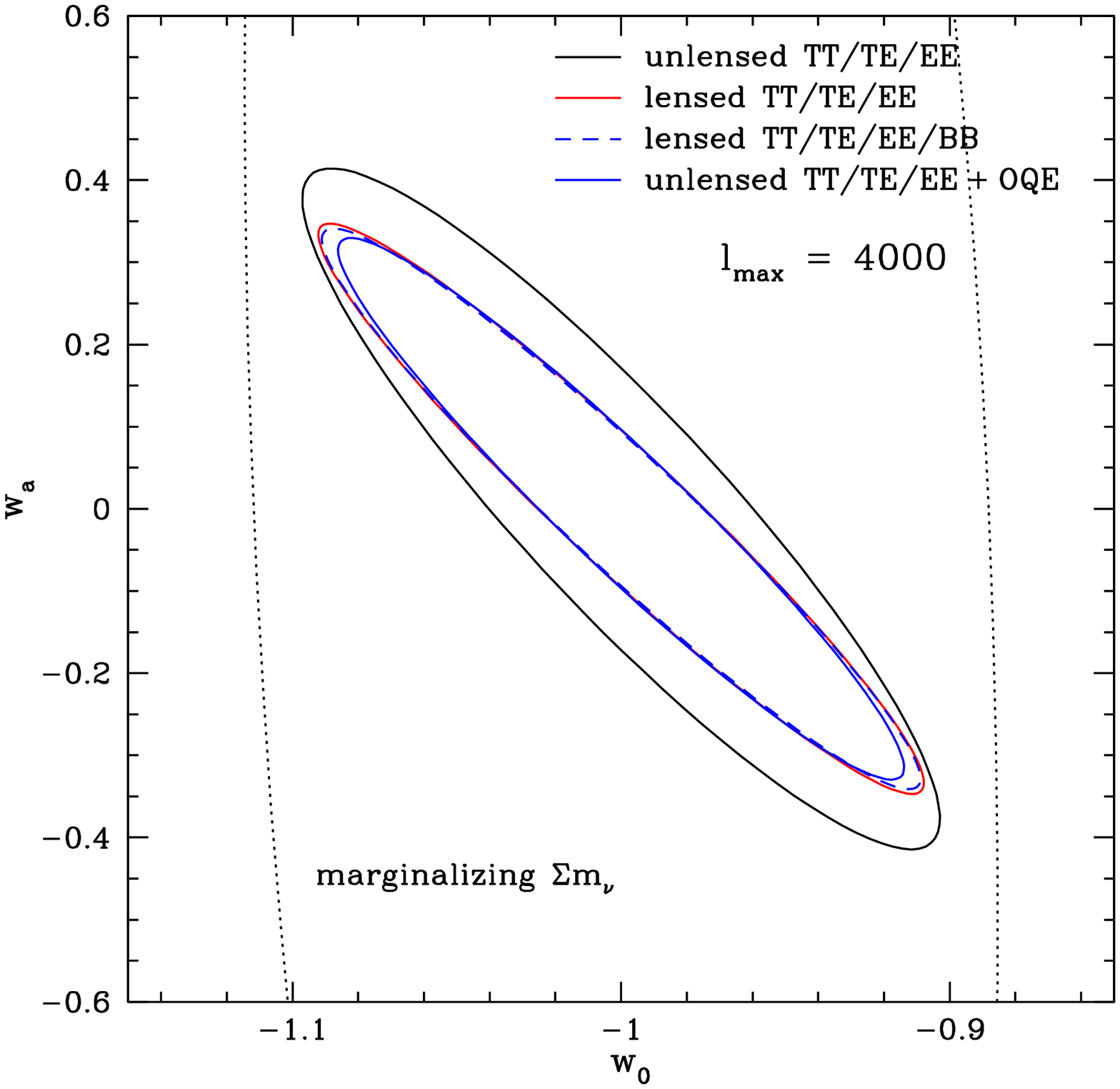}
\end{minipage}
  \caption{As Fig.~\ref{fig:wadata} but here within each panel the
contours correspond to data set types, and the panels use different
systematics levels: $\lmax=1000$ (upper left), 2000 (upper right),
3000 (lower left), 4000 (lower right). Since using lensed TT/EE/TE spectra
is not a matter of simply adding to the Fisher matrix
from unlensed spectra, it is possible for a lensed contour to lie slightly
outside of the unlensed contour, as in the $l_{\rm max}=3000$ case.  
}
\label{fig:walmax}
\end{figure*}

Figure~\ref{fig:walmax} exhibits the analogous situation for different 
systematic limits $\lmax$.  In contrast to the \lcdm\ case, here the 
constraints continue to improve for higher $\lmax$.  There are also 
slight differences between the two methods of fully incorporating 
lensing: use of B-modes or OQE of the lensing potential.  This emphasizes 
that conclusions on systematics or analysis methods should not be based 
solely on examination of the vanilla \lcdm\ cosmology.  Finally, when 
systematics are low, $\lmax=4000$, sufficient information is present in 
the lensed E-modes that further lensing information is unimportant.

\begin{figure}[!htb]
\begin{center}
 \includegraphics*[width=8.63cm]{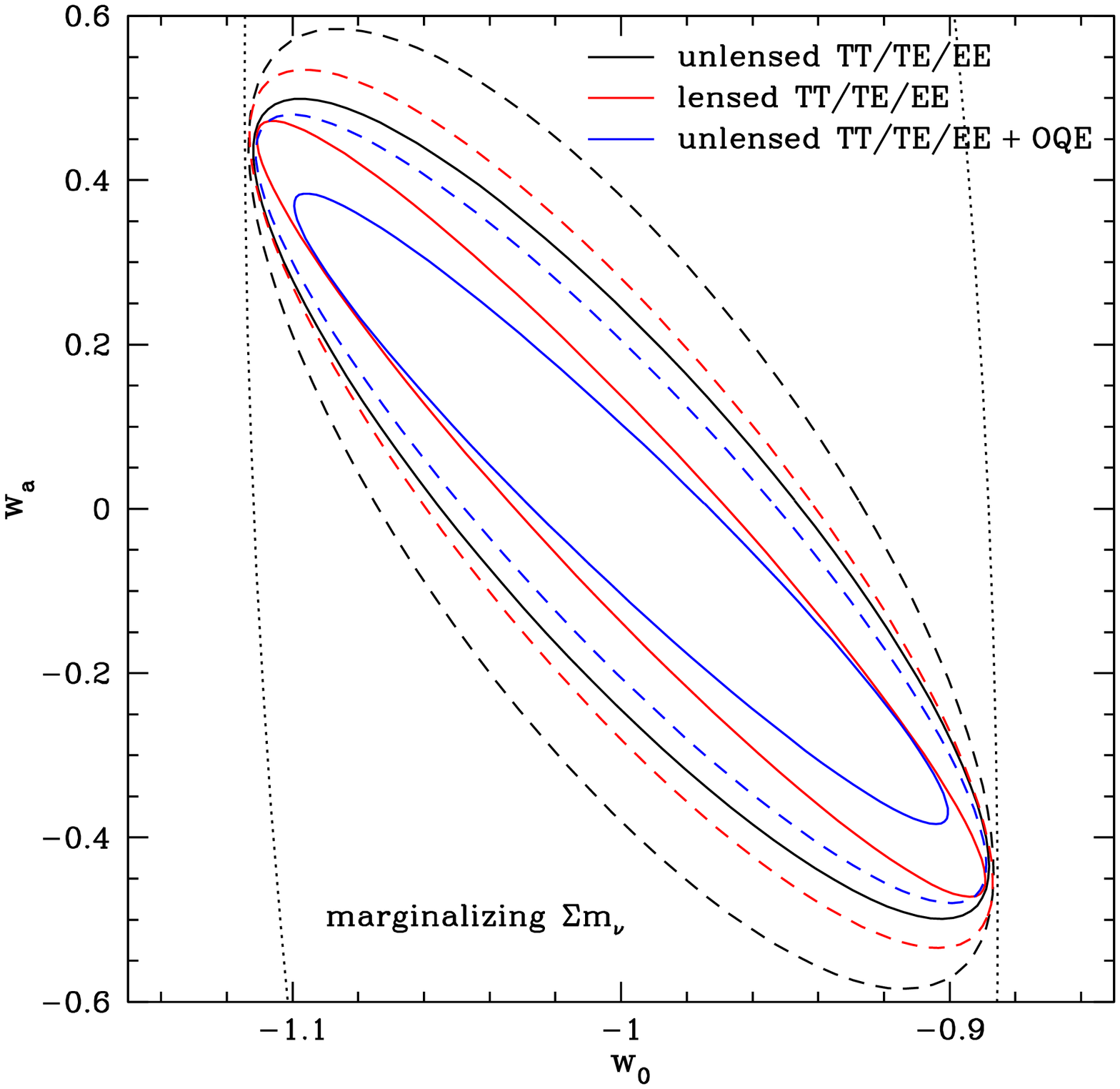}
\caption{Comparing the cosmological constraints on the dark energy 
equation of state parameters from Planck (dashed contours) vs.\ 
CMBpol (solid), taking $\lmax=2000$ and including SNAP-quality 
supernova distances.  The dotted curve gives the constraints from 
supernovae alone. 
}
\label{fig:waplanck}
\end{center}
\end{figure}

The improvements in dark energy estimation that CMB lensing brings is 
illustrated in Fig.~\ref{fig:waplanck} as a function of experiment. 
For Planck, again no lensing information beyond E-modes is useful, though 
the contour area decreases by a factor 1.9 from the unlensed 
case to the OQE case.  By contrast, CMBpol could reduce the likelihood 
contour area by a factor 4.2 relative to the unlensed Planck case, with the 
full lensing information helping by a factor 2.7 relative to unlensed CMBpol.

\begin{figure}[!htb]
\begin{center}
 \includegraphics*[width=8.63cm]{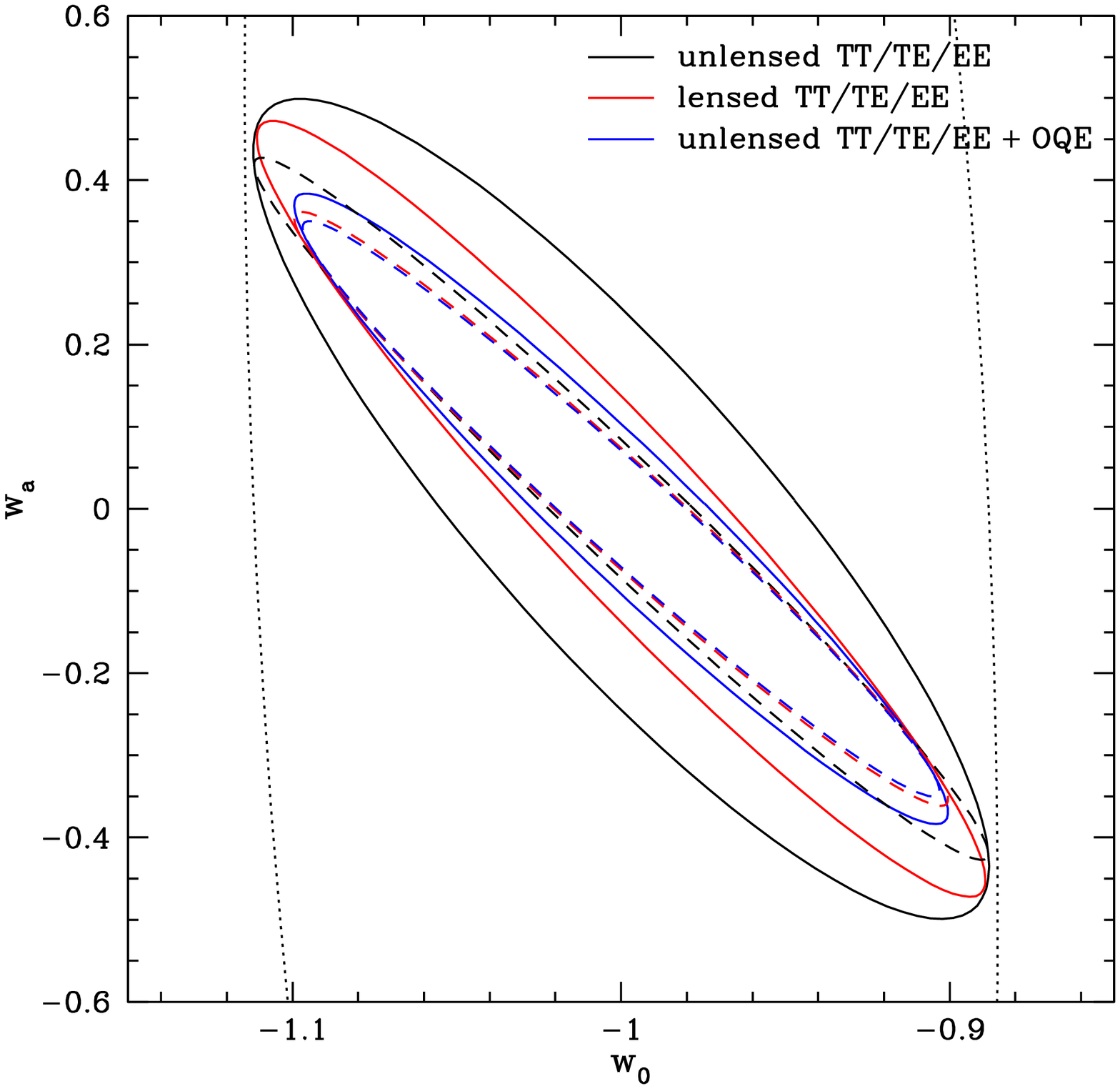}
\caption{ 
As Fig.~\ref{fig:waplanck} for CMBpol only, but here showing 
the effect of fixing $\mnu$ (dashed contours) rather than 
marginalizing over it (solid) as is standard for all parameters not shown. 
}
\label{fig:wanomnu}
\end{center}
\end{figure}

To test the effect of including both neutrino mass and dark energy 
dynamics, Fig.~\ref{fig:wanomnu} shows the likelihood contours for 
the $\lmax=2000$ CMBpol case, marginalizing over vs.\ fixing $\mnu$. 
The fully marginalized uncertainties are 
$\sigma(\mnu)=0.041$, $\sigma(w_0)=0.066$, $\sigma(w_a)=0.25$. 
While the $1\sigma$ limits on the parameters do not change that 
strongly, the total area of the contour is significantly affected. 
For the unlensed (fully lensed) case the area increases by a factor 2.9
(1.5) when properly marginalizing over neutrino mass.  (This 
effect would be more severe when considering CMB data alone.)  
Note that for the CMBpol case the correlation coefficient between 
$w_0$ and $\mnu$ is 0.23 and between $w_a$ and $\mnu$ is $-0.41$; 
while not highly correlated, these are sufficient to give the 
appreciable effect.

\section{Exploring Early Dark Energy \label{sec:ede}} 

In \lcdm, the fractional contribution of dark energy density is of 
order $10^{-9}$ at last scattering.  However, many models exist 
where this can be at the percent level \cite{drw}, with important impacts 
on the sound horizon scale and baryon acoustic oscillations, structure 
formation, and secondary anisotropies 
\cite{drw,doranrob,darkages,sadeh1,linrobb,fllaug08,grossi}. 
Such early dark energy models follow from physics where the dark energy 
traces the energy density of the dominant component of the universe, 
as in high energy physics and string theory models with dilatation 
symmetries \cite{wett88}. 

Although the sound horizon is altered in the presence of early dark 
energy by $\sim(1-\ome)^{1/2}$, this shift can be hidden in the 
temperature power spectrum by compensating changes in the other 
parameters \cite{linrobb}.  This is problematic for baryon acoustic 
oscillation experiments, which use the sound horizon as a standard 
ruler to probe cosmology through distances.  More generally, 
definitive recognition of early dark energy is quite important to 
have confidence in the accurate estimation of the other parameters, 
ensuring that they are not biased due to incorrectly assuming 
no early dark energy.  Furthermore, detection of early dark energy would 
immediately give crucial clues to understanding the nature of dark energy.

Since CMB lensing depends on the growth of structure, it is a good 
candidate for constraining dark energy.  More generally, hints already 
exist in \cite{linrobb} that polarization information can help break 
degeneracies involving early dark energy.  Here we carry out a more 
comprehensive likelihood analysis for unlensed polarization power 
spectra and examine for the first time CMB lensing constraints on 
early dark energy. To do this, we employ the parametrization for the 
fractional dark energy density as a function of scale factor 
proposed by \cite{doranrob},
\beq
\Omega_{de}(a) = \frac{\Omega_{de} - \Omega_e \left(1 - a^{-3 w_0}\right)}{\Omega_{de} + \Omega_m a^{3 w_0}}
+ \Omega_e \left(1 - a^{-3 w_0}\right),
\eeq
where $\Omega_{de}$ is the current dark energy density, 
$\Omega_e$ is the constant dark energy density at early times, 
and $w_0$ is the present dark energy equation of state.  
Hence, the two added parameters $\Omega_e$ and $w_0$ describe the 
dark energy properties. 

\begin{figure*}[!htb]
\begin{minipage}[t]{0.49\textwidth}
\centering
  \includegraphics*[width=\linewidth]{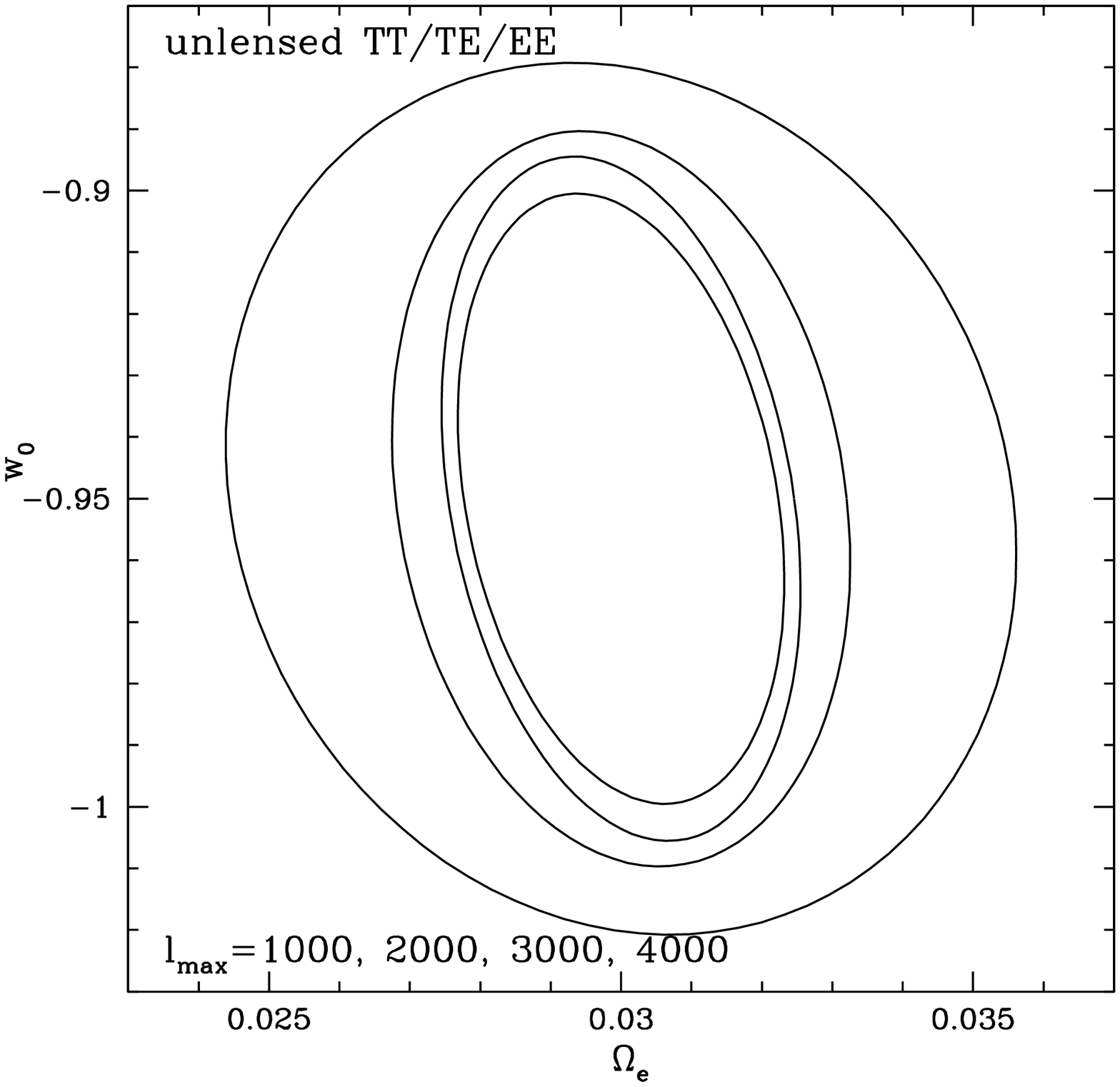}
\includegraphics*[width=\linewidth]{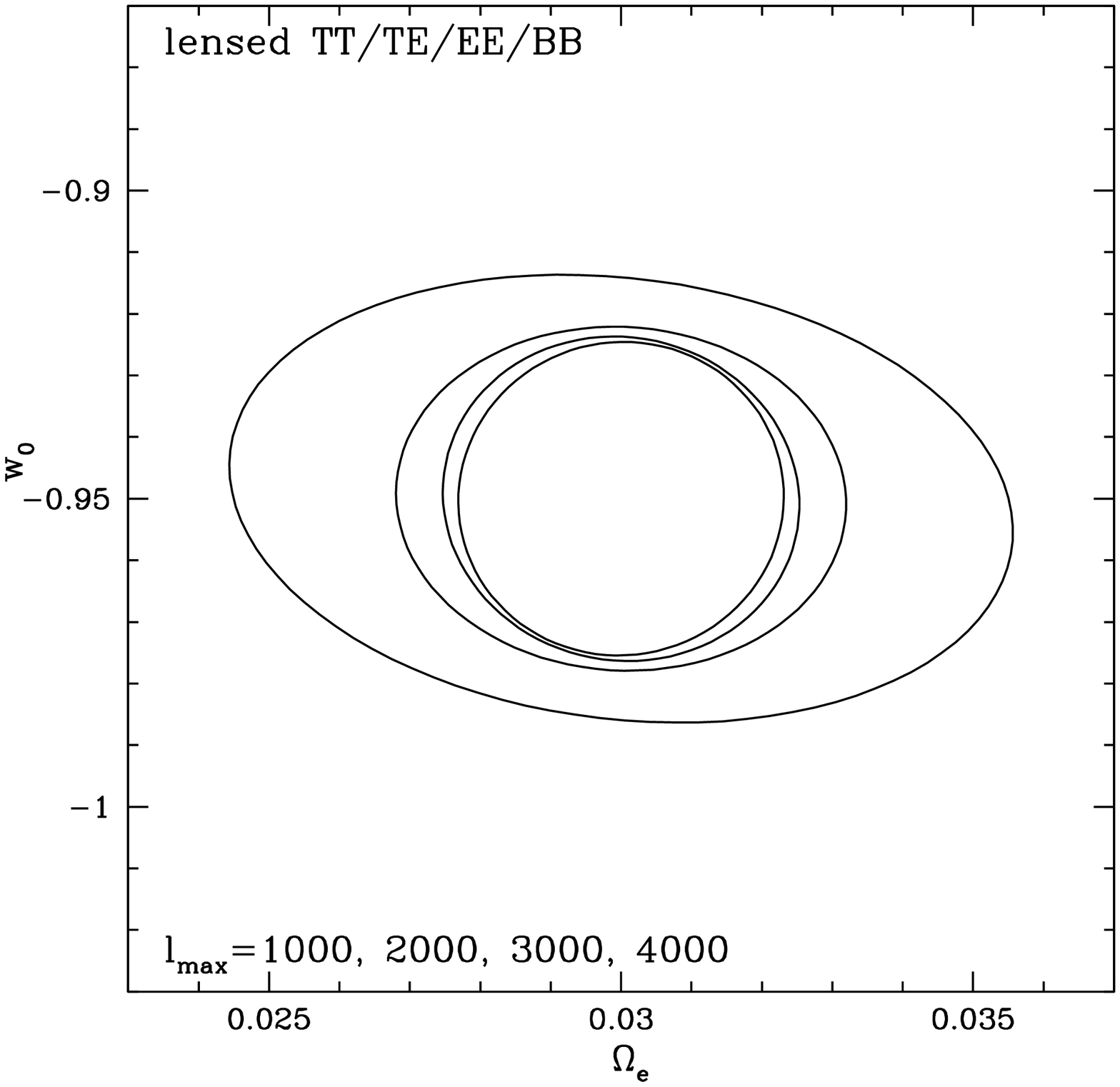}
\end{minipage} \hfill
\begin{minipage}[t]{0.49\textwidth}
\centering
\hfill
  \includegraphics*[width=\linewidth]{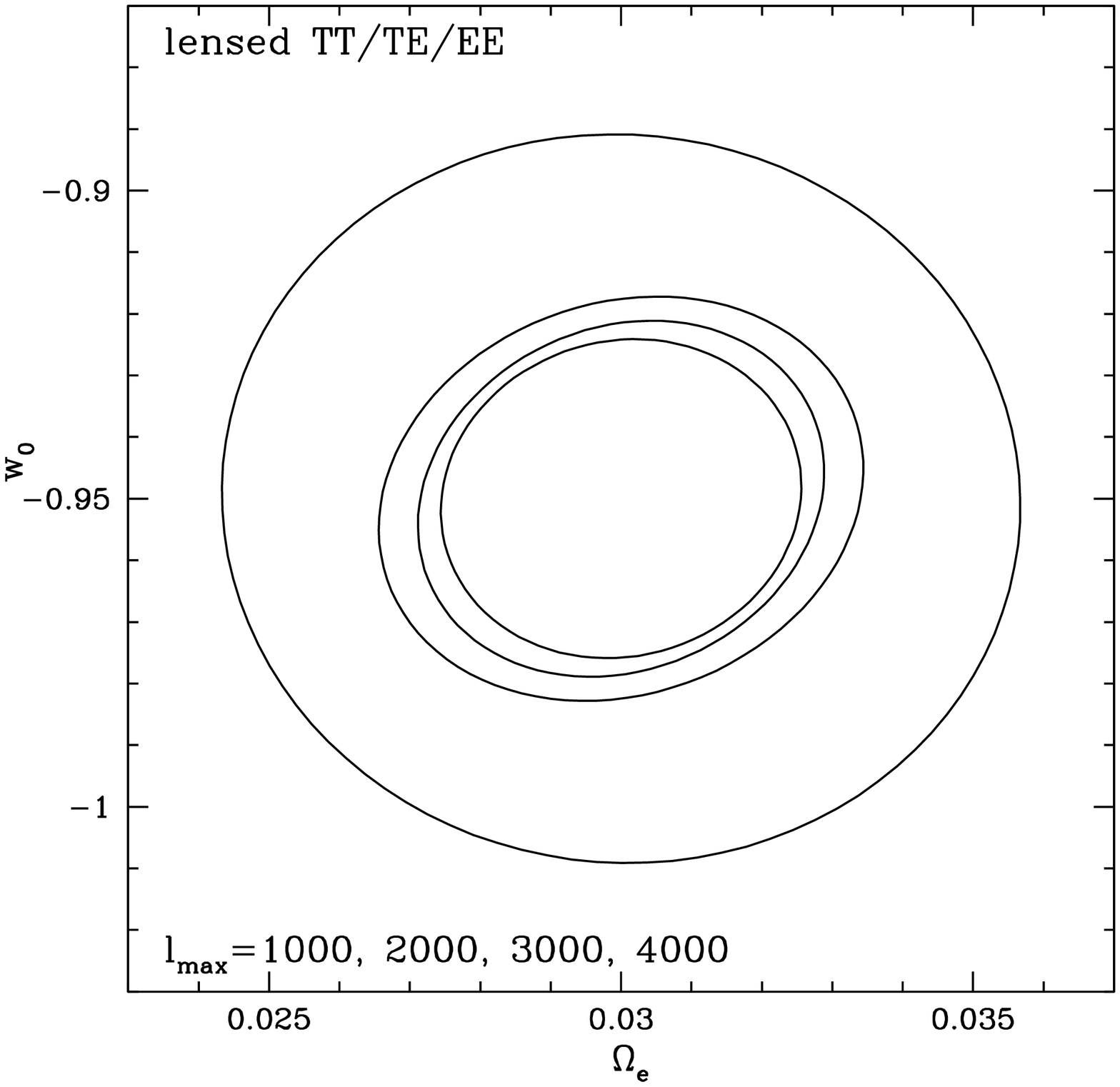} 
\hfill 
  \includegraphics*[width=\linewidth]{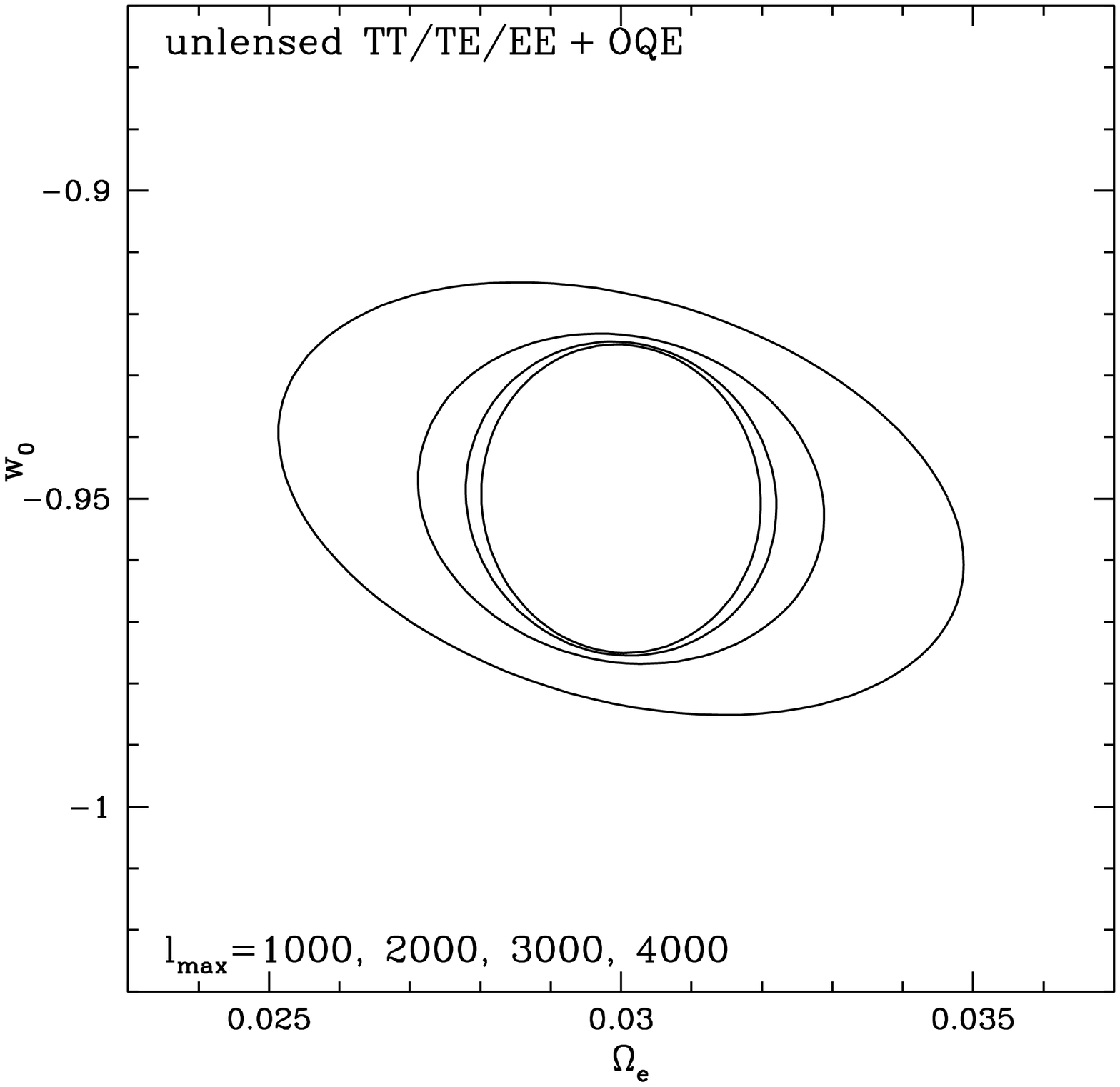}
\end{minipage}
  \caption{Cosmological constraints on the early dark energy 
fraction $\ome$ and present equation of state parameter $w_0$ 
from CMBpol in combination with SNAP-quality supernova distances.  Within
each panel the contours correspond to systematic cuts at $\lmax=1000$,
2000, 3000, 4000 from outer to inner.  The panels use different data cuts:
no lensing (upper left), including lensing from T- and E-modes (upper right),
including lensing from T-, E- and B-modes (lower left), and including lensing 
through the optimal quadratic estimator of the lensing potential (lower right).
}
\label{fig:oedata}
\end{figure*}

Figure~\ref{fig:oedata} shows the constraints in the $w_0$-$\ome$ plane,
marginalizing over the other seven parameters, for different data set 
types.  The fiducial model has $w_0=-0.95$, $\ome=0.03$.  
As in the $w_0$-$w_a$ case, the CMB degeneracies are too strong 
to allow constraints by the CMB alone, so we have again folded in supernova 
distance data (which does not directly constrain $\ome$).  We see that 
unlensed power spectra including polarization information can indeed 
tightly constrain early dark energy.  Adding lensed CMB information in 
fact mostly constrains further $w_0$, having minimal effect on $\ome$. 
Recall from \S\ref{sec:thy} that out to $z\approx2$, the early dark 
energy model looks very much like a standard $w_0$-$w_a$ model that 
would not give appreciable early dark energy density.  Thus, early dark 
energy is too early for even the broad redshift kernel of CMB lensing 
to have significant sensitivity to it.

\begin{figure*}[!htb]
\begin{minipage}[t]{0.49\textwidth}
\centering
  \includegraphics*[width=\linewidth]{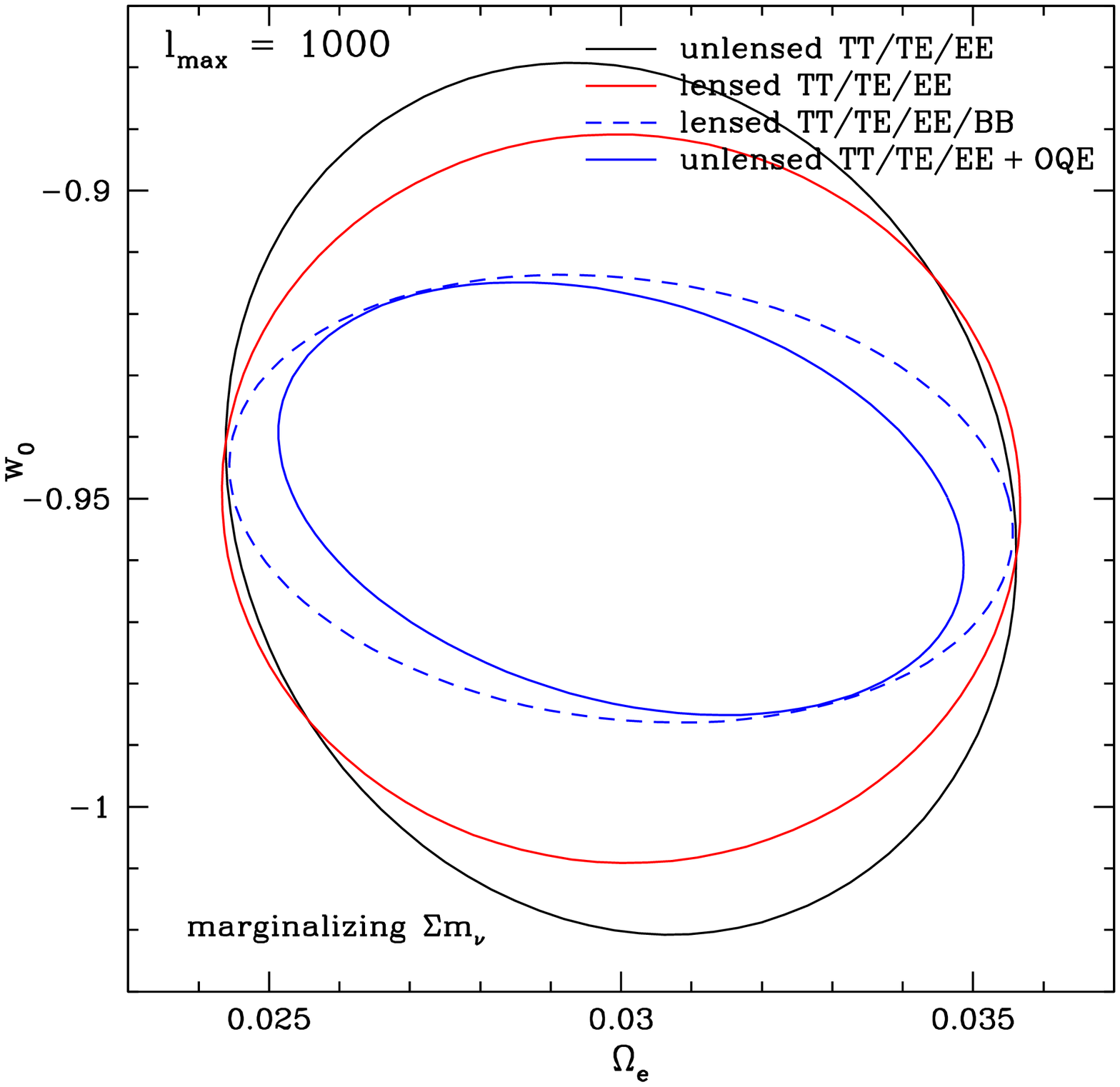}
  \includegraphics*[width=\linewidth]{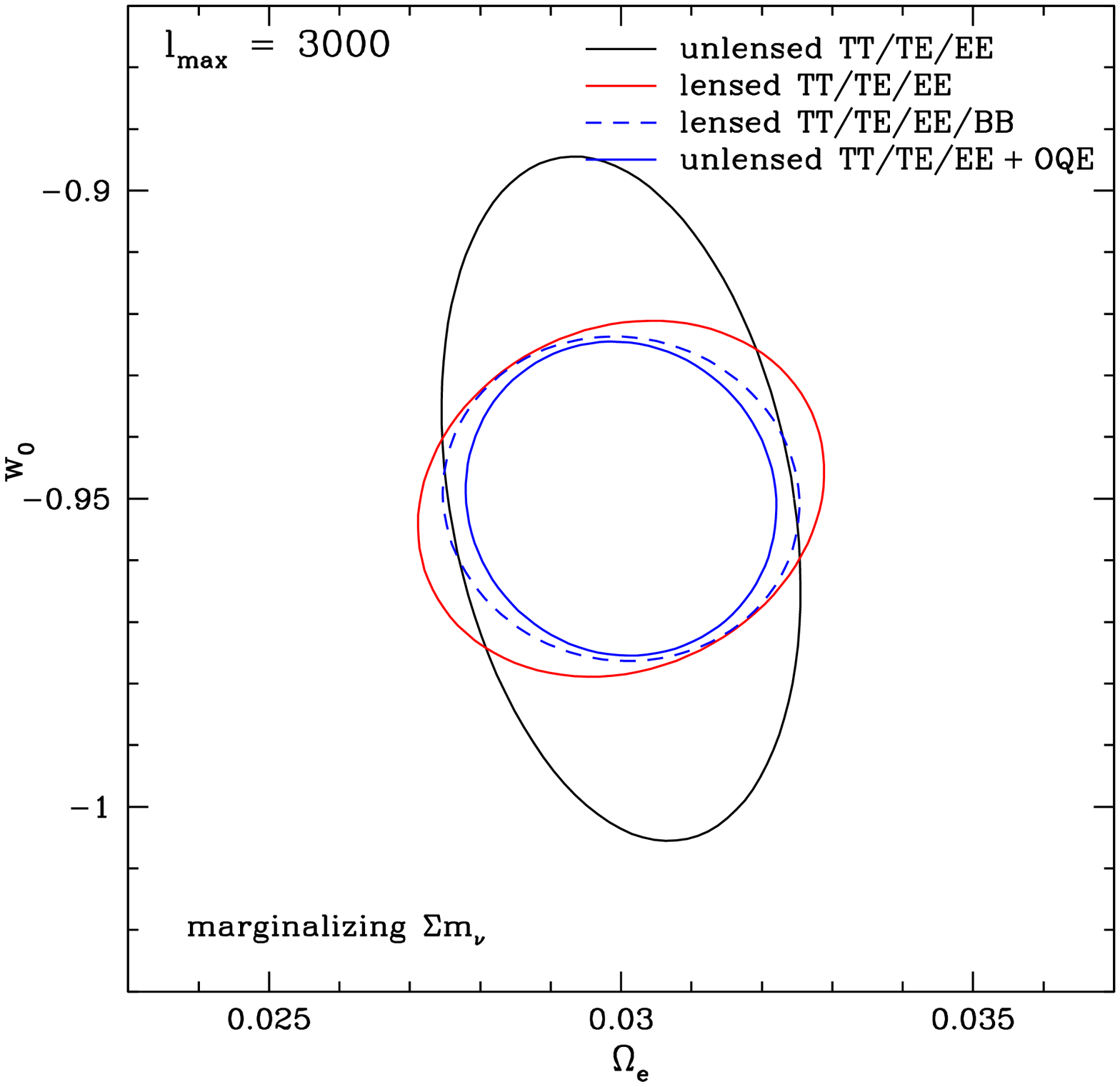}
\end{minipage} \hfill
\begin{minipage}[t]{0.49\textwidth}
\centering
  \includegraphics*[width=\linewidth]{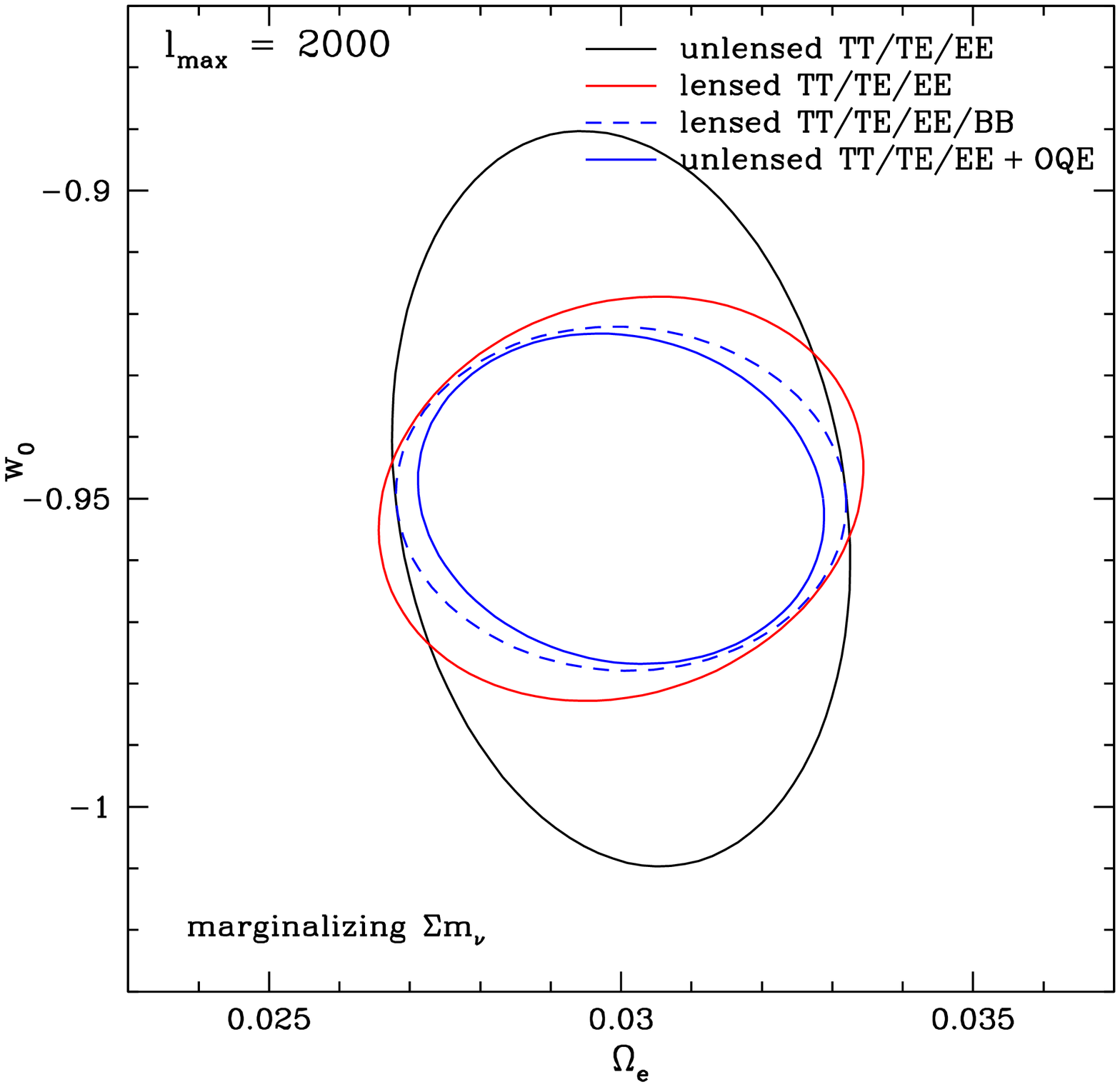}
  \includegraphics*[width=\linewidth]{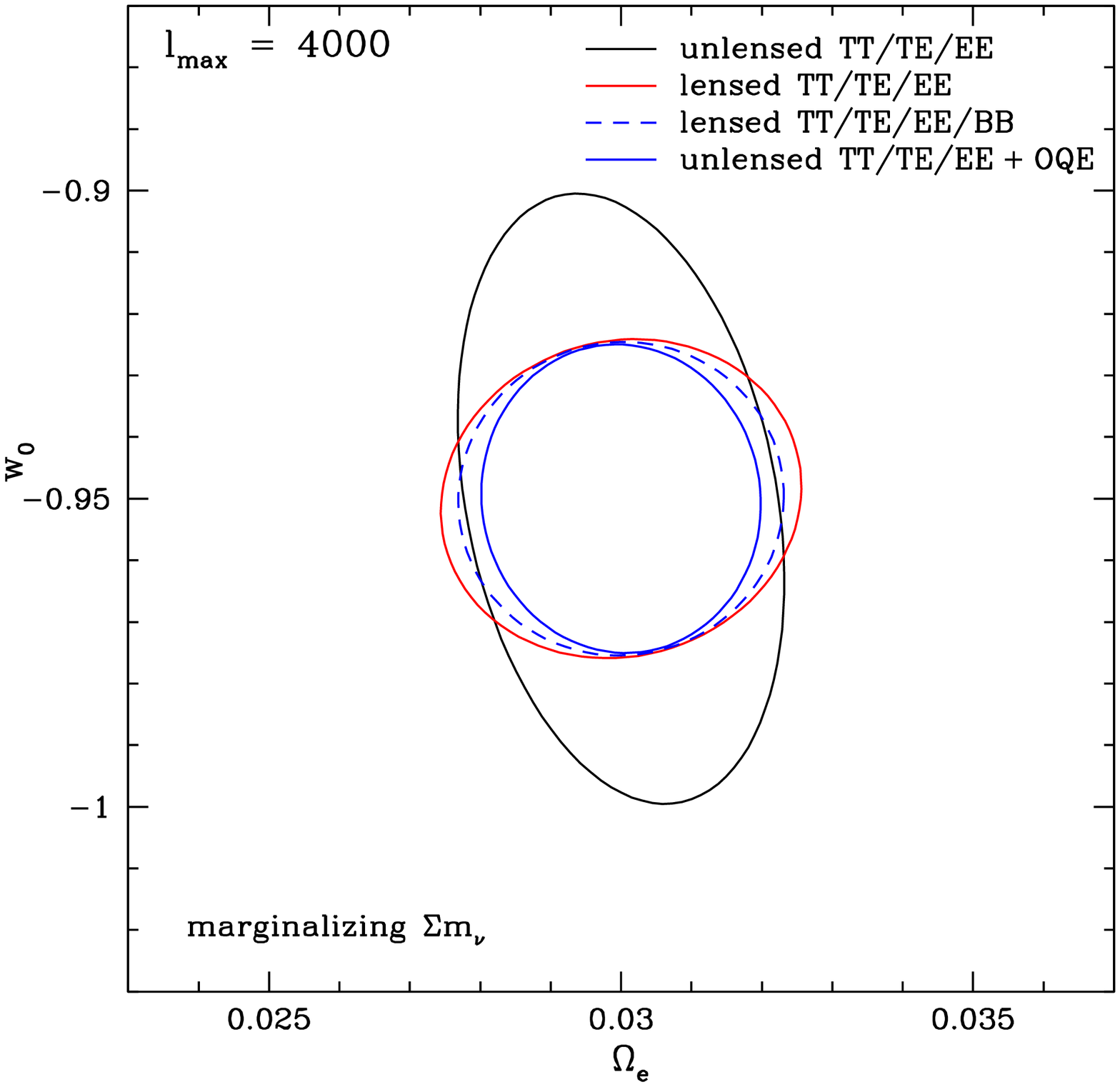}
\end{minipage}
  \caption{As Fig.~\ref{fig:oedata} but here within each panel the
contours correspond to data set types, and the panels use different
systematics levels: $\lmax=1000$ (upper left), 2000 (upper right),
3000 (lower left), 4000 (lower right).
Since using lensed TT/EE/TE spectra
is not a matter of simply adding to the Fisher matrix
from unlensed spectra, it is possible for a lensed contour to lie slightly
outside of the unlensed contour.  
} 
\label{fig:oelmax}
\end{figure*}

Figure~\ref{fig:oelmax} exhibits the analogous situation for different
systematic limits $\lmax$.  Again in contrast to the \lcdm\ case, here 
the constraints continue to improve for higher $\lmax$, although less 
rapidly for $\lmax\gtrsim3000$.  
The fully marginalized uncertainties for the $\lmax=2000$, full lensing 
case are $\sigma(\mnu)=0.047$, $\sigma(w_0)=0.018$, $\sigma(\ome)=0.0019$. 
This is an impressive constraint on the early dark energy density, 
able to give definite guidance to the nature of dark energy, ruling out 
classes of models.

\begin{figure}[!htb]
\begin{center}
 \includegraphics*[width=8.63cm]{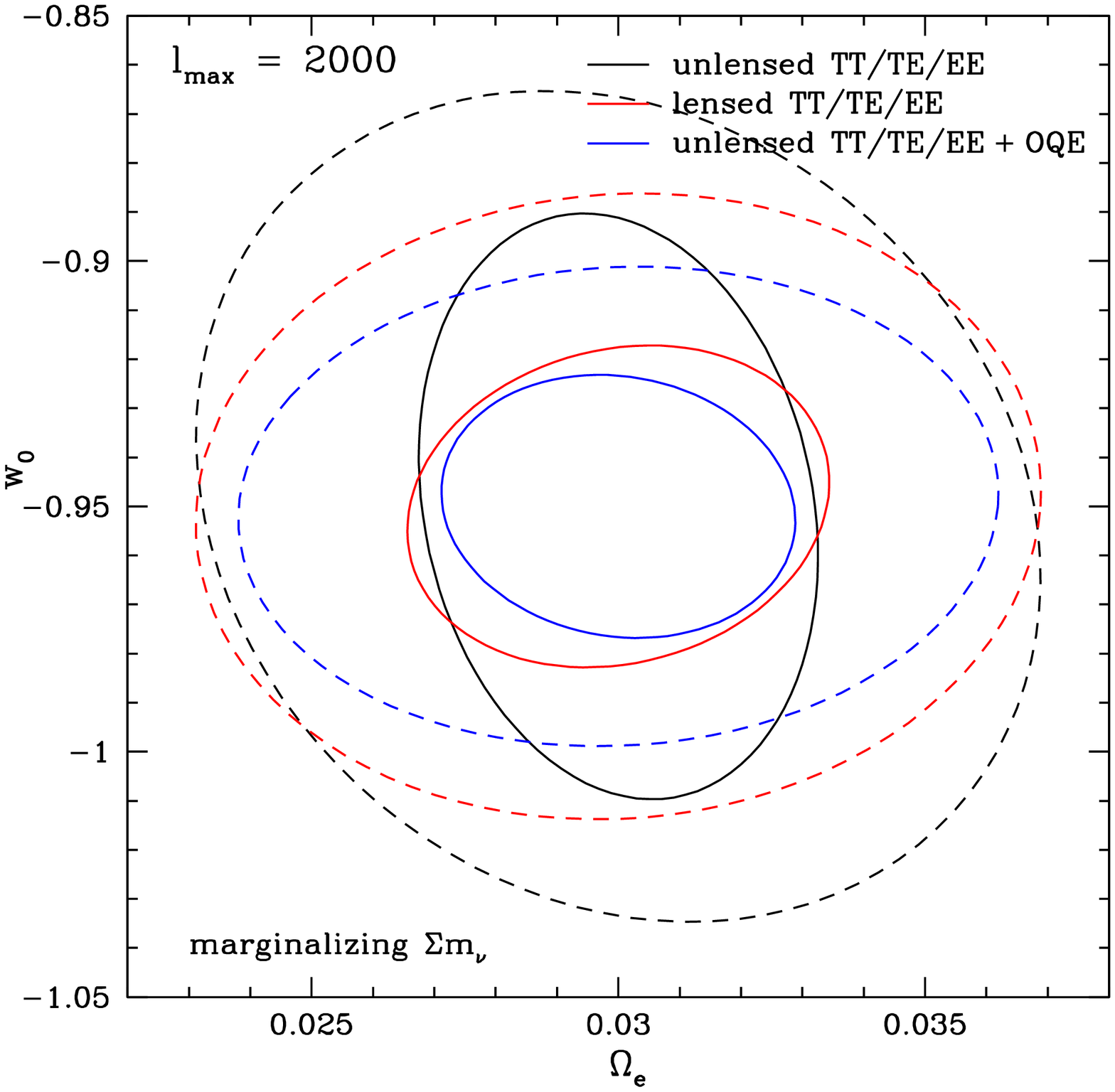}
\caption{Comparing the cosmological constraints on the early dark 
energy fraction $\ome$ and present equation of state parameter $w_0$ from 
Planck (dashed contours) vs.\ CMBpol (solid), taking $\lmax=2000$ 
and including SNAP-quality supernova distances.  
}
\label{fig:oeplanck}
\end{center}
\end{figure}

Because CMBpol would have much better polarization measurements than 
Planck, it will constrain $\ome$ better by a factor 2.2, as shown 
in Fig.~\ref{fig:oeplanck}.  The area of the dark energy properties' 
confidence contour improves by a factor 3.9.

Finally, we summarize our results for the dark energy and neutrino mass 
uncertainties in Table~\ref{tab:sigma_table} for the three cosmological 
models considered, assuming $\lmax=2000$.  However, one should see the 
figures for the full contours.  Due to degeneracies in the presence 
of dynamical dark energy, we add supernova data in these cases to 
constrain the dark energy equation of state, although the uncertainties 
on $\Omega_e$ and $\mnu$ are not strongly affected.

\begin{table}[htbp]
\begin{center}
\begin{tabular*}{0.9\columnwidth}
{@{\extracolsep{\fill}} l l c c c c}
\hline
Model & Experiment & $\sigma(w_0)$ & $\sigma(w_a)$ & $\sigma(\Omega_e)$ &
$\sigma(\Sigma m_{\nu})$ [eV]\\
\hline
$\Lambda$CDM & Planck & -- & -- & -- & 0.11\\
$\Lambda$CDM & CMBpol & -- & -- & -- & 0.036\\
\hline
$w_0$-$w_a$ & Planck+SN & 0.073 & 0.32 & -- & 0.13\\
$w_0$-$w_a$ & CMBpol+SN & 0.066 & 0.25 & -- & 0.041\\
\hline
$w_0$-$\Omega_e$ & Planck+SN & 0.032 & -- & 0.0041 & 0.15\\
$w_0$-$\Omega_e$ & CMBpol+SN & 0.018 & -- & 0.0019 & 0.047\\
\hline
\end{tabular*}
\caption{Uncertainties in parameters beyond standard $\Lambda$CDM for
Planck and CMBpol.  In all cases, we use unlensed temperature and 
polarization spectra and the optimal quadratic estimator 
of the lensing spectrum to extract cosmological information
from the CMB data.  For cases involving dynamical dark energy we fold in 
supernova distance information from a SNAP-like JDEM experiment,  
although this mostly affects only the uncertainties on $w_0$, $w_a$. 
} 
\label{tab:sigma_table}
\end{center}
\end{table}

\section{Shortcut for Joint Dark Energy Constraints \label{sec:jdem}} 

\begin{figure}
\begin{center}{
  \includegraphics*[width=8.6cm]{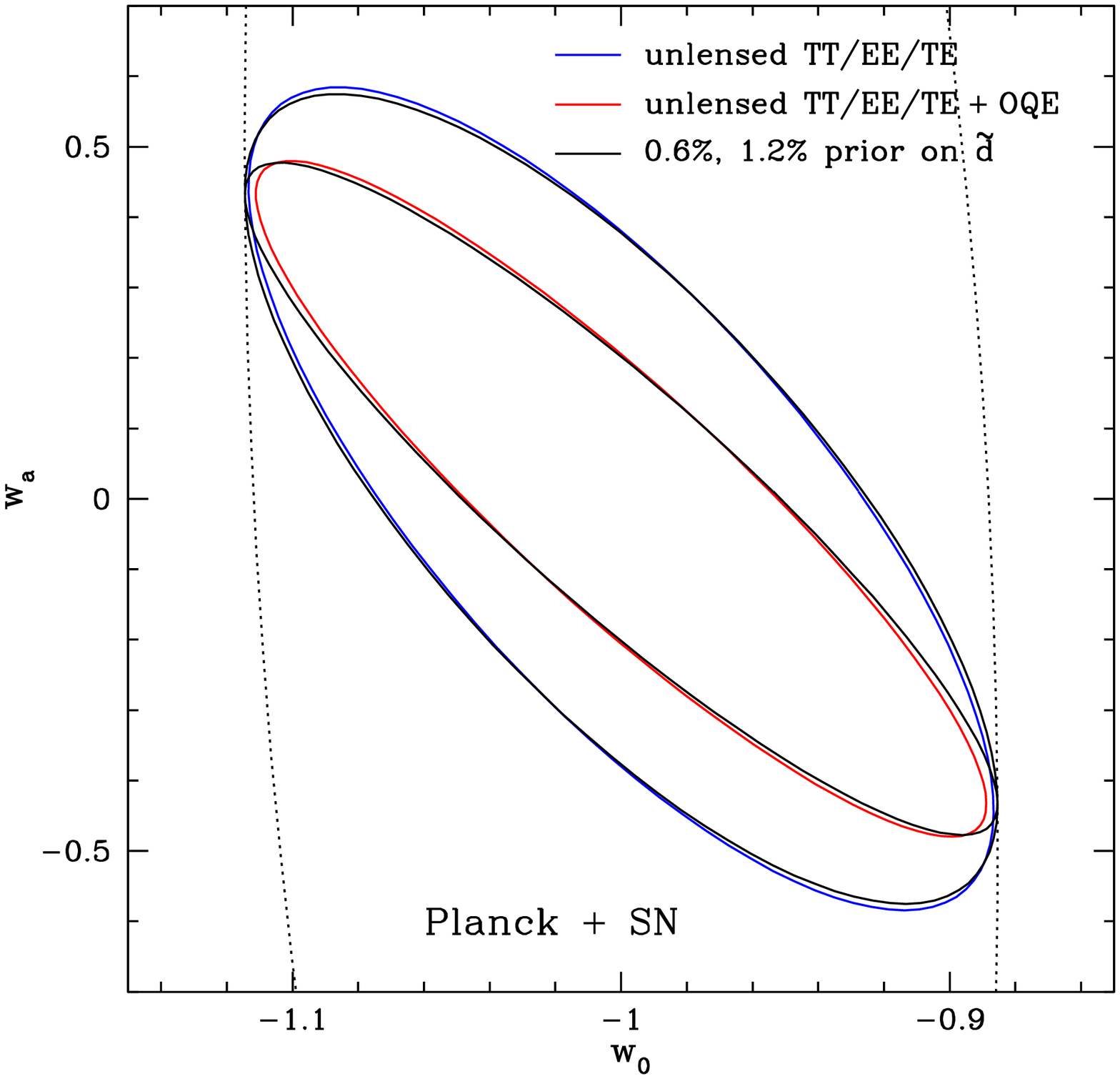}
 \includegraphics*[width=8.6cm]{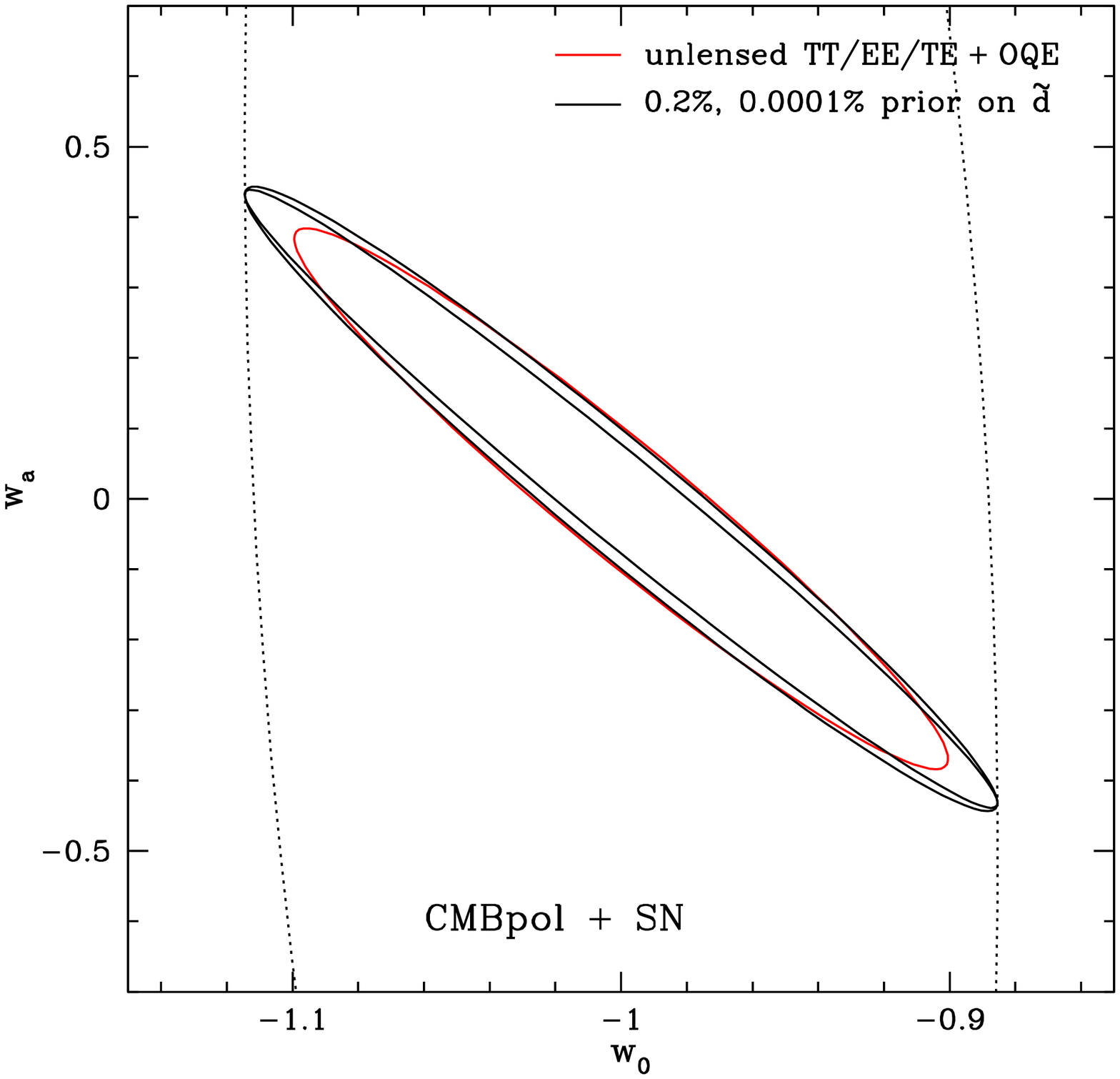}
 }
 \end{center}
\caption{Joint constraints from CMB and supernovae can be well approximated 
by simply replacing the CMB data by an appropriately sized prior on
the shift parameter $\tilde{d} = \sqrt{\omega_m}\, \dlss$.  
[Top panel] Combining Planck data with supernovae, the dark blue 
(light red) curves represent not using (using) lensing information.   
These two cases are well approximated by replacing CMB data by 
$\tilde d$ priors of 1.2\% and 0.6\% respectively. 
[Bottom panel] Combining CMBpol data with supernovae, the light red 
curve represents using lensing information.  This is fairly well 
approximated by a $\tilde d$ prior of 0.2\% (outer black curve).  
Tighter priors have 
little effect (see inner black curve).  Both panels take 
$l_{\rm max}=2000$, and the dotted line in both panels is the 
contour from just supernova data. 
}
\label{fig: shift param}
\end{figure}

As seen in Section~\ref{sec:wa}, when CMB and supernova data are combined, 
we can obtain strong constraints on the nature of dark energy.  
While the supernova data dependence on cosmological parameters is 
straightforward, calculating a CMB Fisher matrix can be quite time 
consuming.  The procedure requires 
computing multiple CMB spectra using a Boltzmann code (CMBeasy in
our case) for different values within a set of cosmological parameters 
in order to obtain the derivatives of the observables with respect to the
cosmological parameters.  

To investigate a range of cosmological models it would therefore be 
quite useful to have a shortcut to calculating 
the constraints on the dark energy parameters
$w_0, w_a, \Omega_{DE}$ from CMB data.  One such shortcut is 
historically well known, 
the shift parameter \cite{BE9807103} to encapsulate the information 
in the temperature power spectrum acoustic peaks.  However, as 
polarization data gets added, other parameters have been suggested 
as additions, e.g.\ the acoustic peak scale $l_A$ \cite{Komatsu:2008hk}, 
although \cite{linrpp} showed that the shift parameter is still quite 
accurate.  Here we investigate the cosmological 
constraints from combining CMB temperature, polarization, and possibly 
deflection, spectra and supernova data, and we show that 
a simple use of the shift parameter has excellent accuracy. 

Specifically, for constraints on the dark energy parameters a strong 
prior on the shift parameter, or reduced distance to last scattering, 
$\tilde{d}=\sqrt{\omega_m}\,\dlss$, is nearly equivalent to the full CMB 
data, even including polarization and lensing data.  That is, 
the CMB Fisher matrix for $\Omega_{DE}, w_0, w_a$
after marginalizing over the other parameters is almost identical
to the Fisher matrix calculated from a single constraint on 
$\tilde{d}\,$\footnote{Note 
this holds for the CMB Fisher matrix itself, without any supernova 
information.}. 
The prior on the quantity $\tilde{d}$ required to match 
the CMB data depends on the CMB experiment and on whether or not we fix 
the neutrino mass.  We emphasize that the level of the prior does not 
correspond to the actual determination of $\tilde d$ from the experiment, 
because the prior also encodes the other spectra information.  
For the CMB experiments we consider, the equivalent prior 
on $\tilde{d}$ is $0.2 \% - 1.2 \%$. 

Note that because early dark energy does not merely affect the projection
of the last scattering surface onto our sky, but also
affects the shape
of the anisotropy spectrum at last scattering directly, we do not expect 
the $\tilde d$ prior to be a complete description there
and indeed we found the prior is not effective in  
this case.

We compare the shift parameter prescription to the use of the actual CMB
Fisher matrix in Fig.~\ref{fig: shift param} by
considering $1\sigma$ joint contours in the $w_0-w_a$ plane
for CMB + SN.  
For Planck (top panel), if we marginalize over
$\sum m_{\nu}$ and if we do not include
the information from CMB lensing, the constraints from the CMB+SN 
are almost exactly the same as those with a $1.2\%$ prior
on $\tilde{d}$.  The constraints are improved quite a bit if the lensing 
information is added. In this case, the constraints are about the 
same as the constraints one gets with a $0.6\%$ prior on $\tilde{d}$.  
In the case of fixing $\sum m_{\nu}$ instead of marginalizing 
over it, the shift parameter prior applies as well, at 0.2\% 
matching the CMB+SN contours whether lensing 
information is used or not. Interestingly, in the case of fixed 
$\sum m_{\nu}$, adding lensing information does not appreciably improve 
the constraints on $w_0$ and $w_a$ any more.  

Note that the Planck experiment can be approximated extremely well by 
the shift parameter prior in all these cases.  The extent $\sigma(w_0)$, 
$\sigma(w_a)$, width $\sigma(w_p)$, area $1/\sqrt{{\rm det}{\bf F}}$, 
and orientation of the dark energy EOS contours match, as seen in 
Fig.~\ref{fig: shift param} and quantified in Table~\ref{tab:dprior}.

\begin{table}[htbp]
\begin{center}
\begin{tabular*}{0.9\columnwidth} 
{@{\extracolsep{\fill}} l c c c c}
\hline 
Data & $\sigma(w_0)$ & $\sigma(w_a)$ & $\sigma(w_p)$ & $\sqrt{{\rm det} {\bf F}}$\\
\hline
SN+Planck & 0.073 & 0.32 & 0.031 & 101\\
SN+0.6\% $\tilde{d}$ & 0.076 & 0.31 & 0.032 & 99\\
\hline
SN+CMBpol & 0.066 & 0.25 & 0.018 & 223\\
SN+0.2\% $\tilde{d}$ & 0.076 & 0.29 & 0.017 & 202\\
\hline 
\end{tabular*}
\caption{Dark energy constraints from supernovae and CMB compared to 
constraints from supernovae and a prior on the shift parameter 
$\tilde{d}$. We assume $l_{\rm max}=2000$ and use the optimal quadratic 
estimator to extract lensing information for both Planck and CMBpol.  
We marginalize over the sum of the neutrino masses and over the other 
parameters of the model.  Note $\sigma(w_p)$ is the width of the 
$w_0$-$w_a$ contour at $w_a=0$ (i.e.\ the uncertainty in constant $w$) 
and $\sqrt{{\rm det} {\bf F}}$ is the inverse area of the contour 
(sometimes used as a figure of merit).} 
\label{tab:dprior} 
\end{center}
\end{table}

For CMBpol (bottom panel of Fig.~\ref{fig: shift param}), the 
constraints on dark energy can be very 
well approximated by a $0.2 \%$ prior on the shift parameter. This is 
true independent of whether one fixes $m_{\nu}$ 
or marginalizes over it because for CMBpol with lensing, 
fixing $m_{\nu}$ only improves the constraints on $w_0$ 
and $w_a$ a little bit compared to marginalizing over $m_{\nu}$ 
(see Fig.~\ref{fig:wanomnu}). Note that making the prior on 
$\tilde{d}$ even smaller than $0.2 \%$ does not change the contour 
significantly. To illustrate this, Fig.~\ref{fig: shift param} shows 
the contour for a prior of $0.0001 \%$, essentially fixing $\tilde d$.  
It is almost the same as the contour for $0.2 \%$.

The combination of supernova data with a prior on $\tilde{d}$ always 
gives an ellipse with ends touching the contour from supernovae alone. 
This means that while both the area enclosed by the contour and the 
uncertainty in $w_a$ may be improved, the uncertainty in $w_0$ is the 
same as the uncertainty from supernova data only. Since Planck 
constraints are described almost perfectly by the shift parameter, 
this is also true for Planck. However, once we include precision 
measurements of polarization by considering CMBpol, the ends of the 
error ellipse can move away from the ``SN only'' contour and thus 
(slightly) improve the constraint on $w_0$. This effect cannot be 
reproduced by the prior on the shift parameter. Hence, for CMBpol, 
the shift parameter prescription works less well than for Planck, 
although it is still quite adequate.  Again, Table~\ref{tab:dprior} 
quantifies the accuracy of substituting the prior in place of the 
full CMB spectra.

\section{Progress in Near-term Experiments: PolarBear \label{sec:inter}} 

In this section we explore the merit of near term ground-based polarization sensitive CMB missions to constrain dark energy and neutrino properties. A number of such experiments are currently being built or have been funded 
including 
BICEP/BICEP2 \cite{bicep}, BRAIN \cite{brain}, C$\ell$OVER \cite{clover}, 
EBEX \cite{Oxley:2005dg}, QUIET \cite{quiet}, 
Spider \cite{spider}, SPTpol.  
Here we focus on one of them, {\sc PolarBear}, as it represents a good combination of the high angular resolution and sensitivity some of these experiments will be capable of. 

\begin{figure*}[!htb]
\begin{minipage}[t]{0.49\textwidth}
\centering
  \includegraphics*[width=\linewidth]{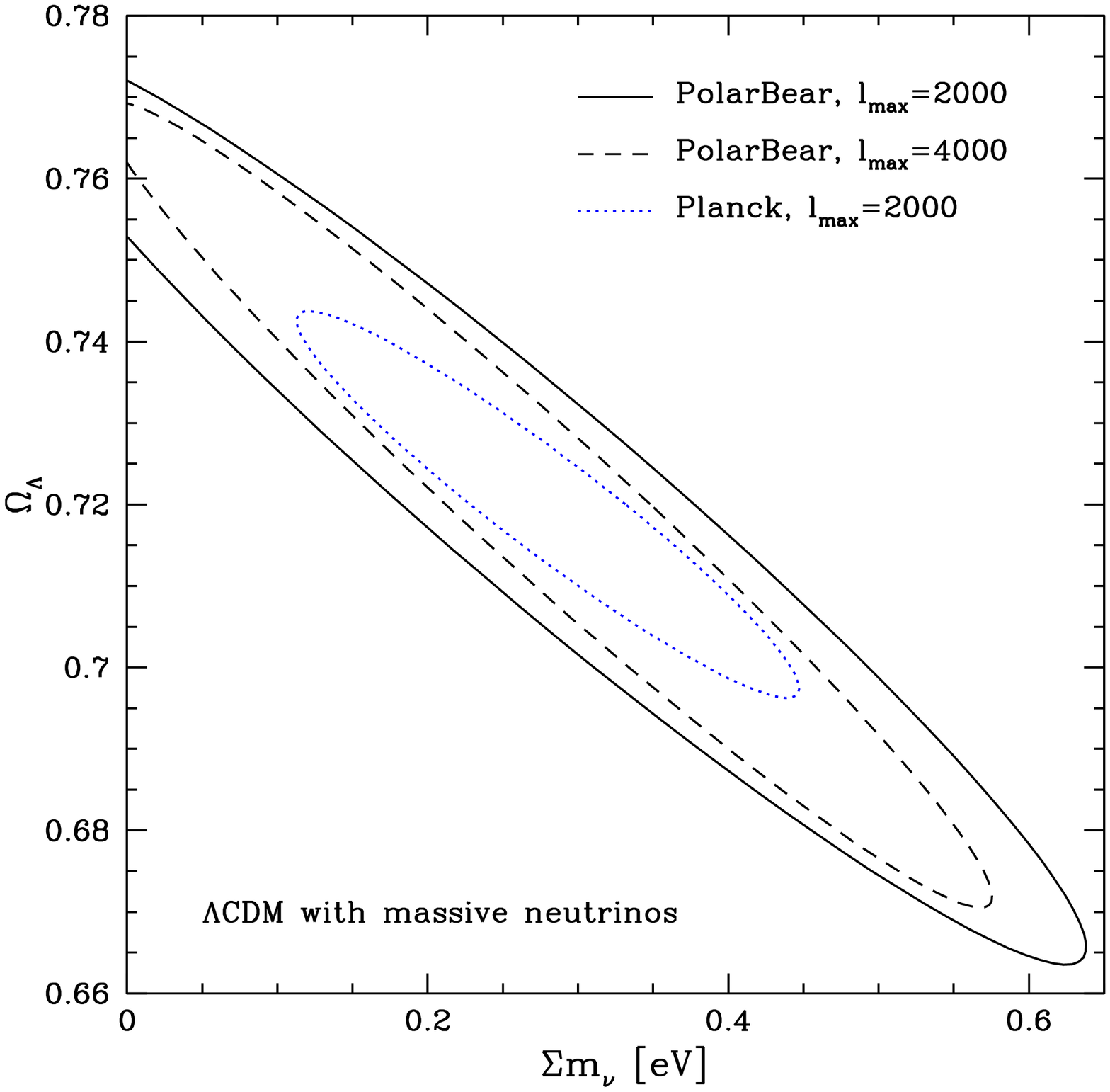}
  \includegraphics*[width=\linewidth]{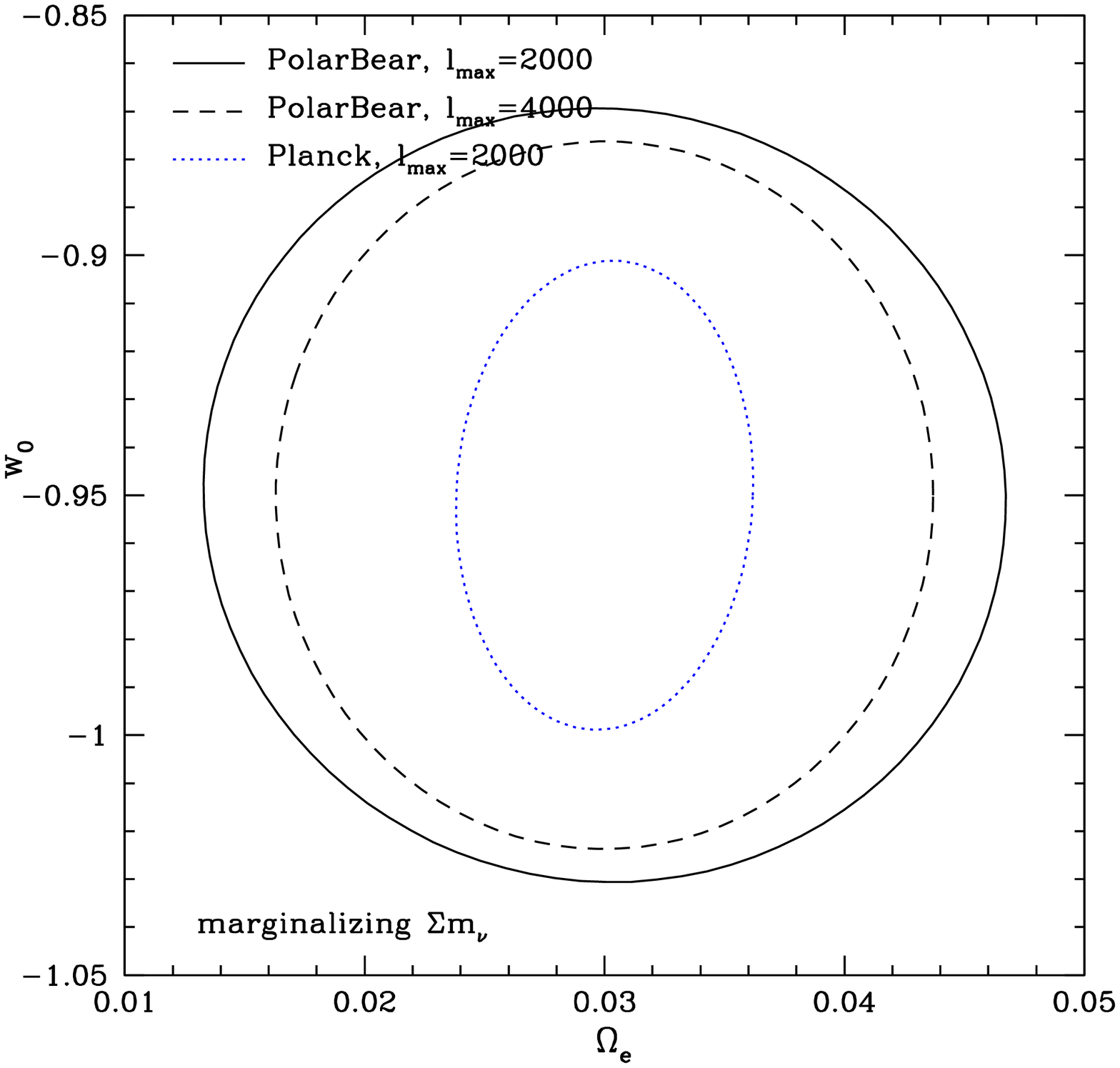}
\end{minipage} \hfill
\begin{minipage}[t]{0.49\textwidth}
\centering
  \includegraphics*[width=\linewidth]{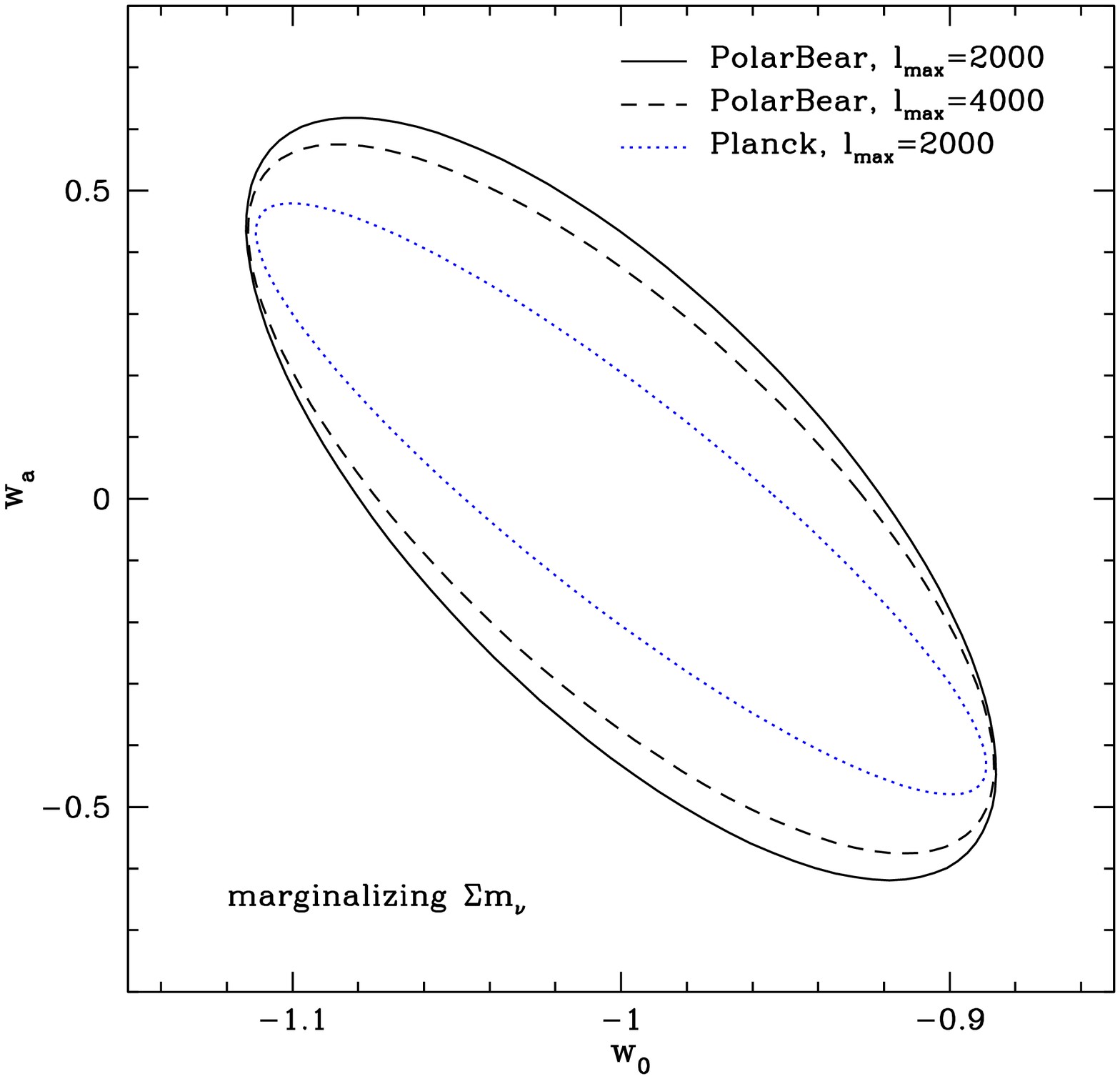}
\end{minipage}
  \caption{Constraints from the intermediate experiment PolarBear are 
not as strong as Planck within the restricted inflationary scenario 
assumed here.  
Contours are constructed using unlensed TT/TE/EE data plus 
the optimal quadratic estimator for the lensing spectrum.  
Blue dotted contours repeat the results for Planck from 
Figs.~\ref{fig:mnuplanck}, \ref{fig:waplanck}, \ref{fig:oeplanck}. 
} 
\label{fig:sptbear}
\end{figure*}

{\sc PolarBear} is a ground based telescope with scheduled beginning of operations in 2009, and deployment to Northern Chile in 2010. It plans to observe 2.5\% of the sky. The low noise of its detectors will enable this experiment to go beyond Planck in imaging the 
B-type polarization pattern, which on small scales is a clear signature 
of gravitational lensing as it cannot be produced by scalar fluctuations. However, the smaller sky coverage does not allow the lensing potential power spectrum to be constrained with as high a signal-to-noise on most scales, making forecasted constraints generally somewhat less good.  To describe this experiment's capabilities, we have adopted specifications from \cite{Tran}, and the 
resulting likelihood contours are shown in Fig.~\ref{fig:sptbear}.

The cosmological constraints from PolarBear lensing reconstruction are less good than those from Planck,
despite the significantly lower noise level. The reason is simply that the limited sky coverage does not allow most modes in 
the temperature, polarization, and lensing potential power spectra to be 
constrained with as high overall signal-to-noise. However the constraints 
are still interesting relative to current limits.  Moreover, we 
particularly note that our parameter space has been limited to not 
include tensor fluctuations, which are a natural consequence of 
inflationary models. With its low noise level PolarBear will attempt to 
measure these gravitational waves from inflation and will help break 
degeneracies between the tensor-to-scalar ratio and other parameters 
that are present in the Planck data.  Furthermore, we have not included 
running of the scalar spectral index; again, PolarBear's high resolution 
and low noise will provide an advantage in breaking degeneracies once 
running is included. 

\begin{figure}[!htb]
\begin{center}
 \includegraphics*[width=8.63cm]{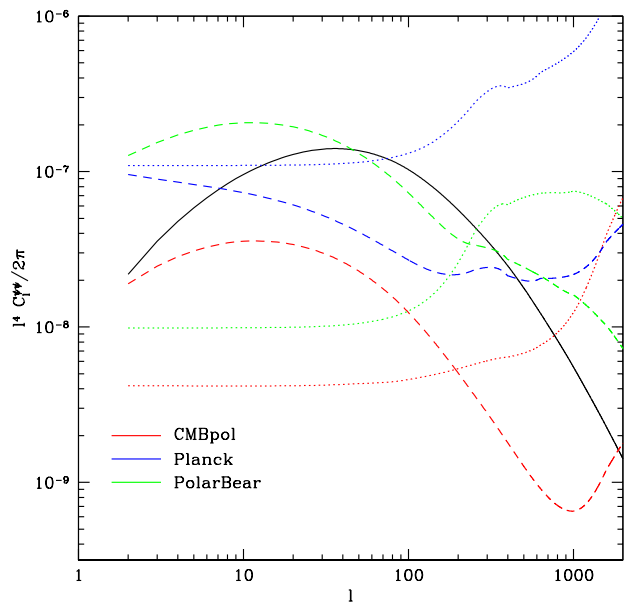}
 \end{center} 
\caption{The lensing potential power spectrum (solid line)
is shown together with the lensing reconstruction noises for the three experiments
considered in this work (dotted lines) and the total error on individual multipoles in the lensing potential,
a combination of sample variance and noise (dashed lines).} 
\label{fig:clpsi} 
\end{figure}

We have found that with Planck the use of the quadratic estimator
vs.\ lensed power spectra leads to a significant improvement of the
constraints on parameters to which lensing is sensitive.
To be specific, we find a $39\%$ improvement on the neutrino mass scale
and a $26\%$ improvement on $\Omega_\Lambda$.  The improvement in the
case of {\sc PolarBear} and {\sc CMBpol} is however only marginal.
To illuminate this trend, in Figure~\ref{fig:clpsi} we plot the power
spectra of the lensing potential and lensing reconstruction noises as
well as the total errors.
The dotted lines show the lensing reconstruction noises for each
experiment.  PolarBear has better capability to map the lensing
potential in the observed patches on the sky than Planck (although
it reconstructs far fewer of these patches and therefore the total
error is larger than for
Planck). The lower lensing noise feeds into the estimation with
the optimal quadratic estimator for reconstruction.

\section{Conclusions \label{sec:concl}}

Continued advances are expected in measuring the cosmic microwave background 
radiation including lower noise and better systematics control, smaller 
beams and wider surveys, and extension to polarization, cross spectra, 
and CMB lensing information.  These will greatly improve our knowledge 
of a variety of cosmological parameters related to primordial 
perturbations.  Here we have explored their impact on physics where the 
CMB has not had as much direct leverage -- extensions to the standard model 
of cosmology such as the necessary neutrino mass and the suspected dynamics 
of dark energy. 

We find the following general points to guide the design and analysis 
of CMB experiments, both ground based and the CMBpol satellite concept:  

\begin{itemize} 

\item Systematics, such as point sources and other foreground 
contamination, will affect the lensing potential and other power spectra, 
and should be removed at the level of at least $\lmax=2000$.  
Constraints improve only slowly for higher $\lmax$ when using the full 
information in the CMB. 

\item Analysis of gravitational lensing of the CMB can proceed 
either through consideration of induced B-mode polarization or 
through an optimal quadratic estimator directly of the deflection 
field; the optimum is not steep so the two approaches are nearly 
equivalent for these purposes with data beyond Planck. 

\item For exploration of suites of cosmological models, we establish 
the accuracy of a shortcut in terms of an effective prior on the 
CMB shift parameter.  This is remarkably efficient in summarizing 
the information from the CMB spectra. 

\end{itemize} 

Determination of the sum of neutrino masses can be accomplished 
by CMBpol with an uncertainty of 0.05 eV, marginalizing over all other 
parameters including dark energy properties.  This corresponds to 
greater than a $5\sigma$ detection for the fiducial value adopted, 
and represents a factor 3 improvement over Planck expectations. 
Restricted to a \lcdm\ cosmology, the constraints tighten by a 
factor $\sim1.3$.  

Determination of the dynamical properties of dark energy is less powerful. 
Complementary information, such as from distance measurements, is 
required with the leverage of the two data sets together allowing 
significant constraints.  The present dark energy equation of state 
$w_0$ could be estimated to 0.07 and the time variation $w_a$ to 
0.25, including marginalization over other cosmological parameters 
including neutrino mass.  This would improve further as other 
probes are added.  While the marginalized constraints do not improve 
greatly in going from Planck to CMBpol, the area of the uncertainty 
contour shrinks by a factor 2. 

The most significant impact from the CMB comes within early dark energy 
models.  Here the improvement from Planck to CMBpol is a factor 2 in 
estimation of both $w_0$ (to 0.02 for CMBpol plus distances) and early 
dark energy density $\ome$ (to 0.002 for CMBpol plus distances), while 
the uncertainty area shrinks by a factor 4.  This provides the 
possibility of a $\sim10\sigma$ detection of early dark energy, which 
would immediately revolutionize our physics thinking.  

CMB lensing offers an intriguing new window on the universe, especially 
because of its sensitivity to the properties of expansion and growth 
in the poorly probed epoch $z\approx1-4$.  
Experiments nearly in 
the process of data collection will teach us not only about the primordial 
conditions but also about the interesting period when dark energy first 
becomes significant, as well as establishing a link to terrestrial 
experiments to measure the neutrino masses.

\acknowledgments 

We thank Georg Robbers for tireless advice on \ceasy, and also thank 
Wayne Hu and Sudeep Das for useful exchanges.  
This work was supported in part by the Director, Office of Science, 
Office of High Energy Physics, of the U.S.\ Department of Energy under 
Contract No.\ DE-AC02-05CH11231. OZ acknowledges funding by the 
Berkeley Center for Cosmological Physics.

{}

\end{document}